\documentclass[useAMS,usegraphicx,usenatbib]{mn2e}
\usepackage{color}

\def\lesssim{\,\lower2truept\hbox{${<\atop\hbox{\raise4truept\hbox{$\sim$}}}$}\,}
\def\gtrsim{\,\lower2truept\hbox{${>\atop\hbox{\raise4truept\hbox{$\sim$}}}$}\,}

\def\rev{}

\newcommand{\Kpc}{~h^{-1}~{\rm kpc}}

\title[BCGs in simulations]{Brightest cluster galaxies in cosmological simulations: achievements and limitations of AGN feedback models}

\author[Ragone-Figueroa et al.]{
\parbox[t]{\textwidth}{
Cinthia Ragone-Figueroa$^{1,2}$\thanks{Email: cin@oac.uncor.edu},
Gian~Luigi Granato$^{2}$\thanks{Email: granato@oats.inaf.it},
Giuseppe Murante$^{2}$,
Stefano Borgani$^{3,2}$
and Weiguang Cui$^{3,2}$
}
\vspace*{6pt} \\
  $^1$ Instituto de Astronom\'ia Te\'orica y Experimental (IATE),\\
  Consejo Nacional de Investigaciones Cient\'ificas y T\'ecnicas de la Rep\'ublica Argentina (CONICET),\\ Observatorio
  Astron\'omico, Universidad Nacional de C\'ordoba, Laprida 854, X5000BGR, C\'ordoba, Argentina\\
  $^2$ Istituto Nazionale di Astrofisica INAF, Osservatorio Astronomico di
  Trieste, Via
  Tiepolo
  11, I-34131
  Trieste, Italy \\
  $^3$ Astronomy Unit, Department of Physics, University of Trieste,
  via Tiepolo 11, I-34131 Trieste, Italy\\
}

\begin{document}

\newcommand{\cinthia}[1]{{\bf\textcolor{red}{ #1}}}
\date{Accepted 5 September 2013; Received in original form 9 August 2013}

\maketitle

\begin{abstract}
We analyze the basic properties of Brightest Cluster Galaxies (BCGs) produced
by state of the art cosmological zoom-in hydrodynamical simulations. These
simulations have been run with different sub-grid physics included. Here we
focus on the results obtained with and without the inclusion of the
prescriptions for supermassive black hole (SMBH) growth and of the ensuing
Active Galactic Nuclei (AGN) feedback. The latter process goes in the right
direction of decreasing significantly the overall formation of stars. However,
BCGs end up still containing too much stellar mass, a problem that increases
with halo mass, and having an unsatisfactory structure. This is in the sense
that their effective radii are too large, and that their density profiles
feature a flattening on scales much larger than observed. We also find that our
model of thermal AGN feedback has very little effect on the stellar velocity
dispersions, which turn out to be very large. Taken together, these problems,
which to some extent can be recognized also in other numerical studies
typically dealing with smaller halo masses, indicate that on one hand present
day sub-resolution models of AGN feedback are not effective enough in
diminishing the global formation of stars in the most massive galaxies, but on
the other hand they are relatively too effective in their centers. It is likely
that a form of feedback generating large scale gas outflows from BCGs
precursors, and a more widespread effect over the galaxy volume, can alleviate
these difficulties.
\end{abstract}


\section{Introduction}
\label{sec:intro}

Brightest Cluster Galaxies (BCGs) are an important test for our understanding
of the physical processes driving galaxy evolution, specifically at the massive
end of the mass distribution of galaxies where they dominate, and within the
densest environments they inhabit. Establishing to what extent they differ from
the general (massive) early type galaxies population, is an active topic of
research. From the very fact that they form and live in the center of galaxy
clusters, it is natural to expect differences. Indeed, Bernardi (2009),
studying large samples of such galaxies drawn from the Sloan Digital Sky Survey
(SDSS), concluded that they are larger, at fixed stellar mass, than the general
early type population and their velocity dispersion increases less with stellar
mass. On the other hand, for the few BCGs with available estimates of the
central SMBH mass, no deviations from the general $M_{BH}-M_{bulge}$ or
$M_{BH}-\sigma$ are apparent (Graham \& Scott 2013; McConnel \& Ma 2013; {\rev
but see e.g.\ Hlavacek-Larrondo et al.\ 2012 for indirect claims of
ultramassive SMBHs in BCGs in strong cool core clusters}).

For a long time, significant and persistent tensions have been repeatedly reported
between theoretical predictions of models of galaxy formation, and
observations. These problems are most likely related to the well known
difficulty, arising in the attempt to use galaxy formation to test the
cosmological paradigm, that many processes driving galaxy evolution (notably
star formation and feedback) are poorly understood from a theoretical point of
view, and occur far below the resolution of cosmological simulations. Thus they
are included by means of very approximate and uncertain sub-grid prescriptions.
As a result, quoting the not so surprising outcomes of the Aquila comparison
project (Scannapieco et al.\ 2012), wherein thirteen cosmological gasdynamical
codes evolving identical initial conditions have been considered,
 ``state-of-the-art simulations cannot yet uniquely predict the properties of the
baryonic component of a galaxy, even when the assembly history of its host halo
is fully specified''. Despite the many efforts in the field, that lead to
substantial progresses, many problems still remain.

As for high mass galaxies, and in particular for BCGs, the challenge is to
avoid excessive cooling and star formation, in particular at late time.\footnote{A related,
but still debated issue, is that also the mass growth of BCGs by dry mergers at $z
\lesssim 1$  could be over-predicted by semi analytical models including AGN
feedback (e.g.\ Stott et al.\ 2010, 2011; Lidman et al.\ 2012; Lin et al.\
2013)}  Supernovae feedback, invoked in computations to counteract the efficient cooling
and star formation in low mass haloes, cannot do the same job at high masses.
Actually, it has the side effect of leaving too much gas available for
accretion into massive haloes, wherein it forms over-massive galaxies at
relatively low redshift (see e.g.\ Benson 2010 and references therein for more
details). At present, the most promising solution for this high mass
overcooling problem is the (negative) feedback arising from AGN activity. Somewhat
surprisingly, this physical process has been ignored in computations for a long time.
However, starting from about a decade ago, it has progressively included in most models,
both semi-analytic ones as well as simulations (e.g.\ Granato et al.\ 2004;
Springel et al.\ 2005; Croton et al.\ 2006; Bower et al.\ 2006; Monaco et al.\
2007; Sijacki et al.\ 2007; Somerville et al.\ 2008; McCarthy et al.\ 2010;
Fabjan et al. 2010; Martizzi et al. 2012a).


To test the role of the various sub-resolution physical processes in
cosmological settings, we have undertaken a program of zoomed-in simulations of
a quite large sample of cluster-sized objects. These simulations have been
carried out by incrementally increasing the physics considered, and have been
already exploited for several purposes (see Planelles et al.\ 2013 and
references therein). The main aim of the present study is to compare the basic
properties of the BCGs, as predicted by the simulations wherein all considered
physical processes are included, as well as by those wherein only AGN feedback
is switched off, with those of real BCGs. The properties we will consider are
total stellar masses, sizes, stellar velocity dispersions and density profiles.
This is to clarify the potential importance of the AGN phenomena in affecting
galaxy formation at the highest mass end, and to assess to what extent the
current implementation of the process is satisfactory.

The main difference with respect to previous cosmological simulations, in which
the effect of the inclusion of AGN feedback on BCGs has been considered (e.g.\
Sijacki et al.\ 2007; McCarthy et al.\ 2010; Puchwein et al.\ 2010; Martizzi et
al.\ 2012a,b; Dubois et al.\ 2013), is that our simulations extend to larger
cluster masses, {\rev and/or} allow the selection of larger samples of BCGs.
Thus, this is the first study in which the effectiveness of AGN feedback model
in limiting the mass growth of BCGs, is extensively tested on the scales of
rich galaxy clusters.

In the following all quantities are computed for $h=0.72$, the value adopted in
the simulations, unless otherwise specified.

The paper is organized as follow: in Section \ref{sec:method} we describe the
simulations analyzed in this paper, and the post-processing performed to
generate mock images of the clusters. In Section \ref{sec:results} we present
and describe our results, which are then summarized and discussed in Section
\ref{sec:discussion}. The Appendix is devoted to a more detailed description of
the sub-grid recipes used to treat AGN feedback.

\section{method}
\label{sec:method}
\subsection{The simulated clusters} \label{simulation}
The clusters are extracted from high resolution re-simulations of 29 Lagrangian
regions, taken from a low resolution N-body cosmological simulation. The parent
simulation follows $1024^3$ DM particles within a periodic box of co-moving
size $1 h^{-1} Gpc$, assuming a flat $\Lambda$CDM cosmology: matter density
parameter $\Omega_m = 0.24$; baryon density parameter $\Omega_b = 0.04$; Hubble
constant $h = 0.72$; normalization of the power spectrum  $\sigma_8 = 0.8$;
primordial power spectral index $n_s = 0.96$. Each Lagrangian region has been
re-simulated at higher resolution employing the {\it Zoomed Initial Conditions}
technique (Tormen, Bouchet \& White, 1997). The Lagrangian regions are large
enough to ensure that within five virial-radii of the central cluster only high
resolution particles are present.

The simulations have been run using the TreePM-SPH {\small GADGET-3} code, an
improved version of {\small GADGET-2} (Springel 2005), and adopting several
levels of complexity for the physical processes involved. In the high
resolution region gravitational force is computed by adopting a
Plummer-equivalent softening length of $\epsilon = 5 h^{-1}$ kpc in physical
units below z = 2, while being kept fixed in comoving units at higher redshift.
As for the hydrodynamic forces, we assume the minimum value attainable by the
Smoothed Particle Hydrodynamics (SPH) smoothing length of the B-spline kernel
to be half of the corresponding value of the gravitational softening length. In
the high resolution region, the mass of DM particles is $8.47 \times 10^8
h^{-1} \mbox{M}_\odot$, and the initial mass of each gas particle is $1.53
\times 10^8 h^{-1} \mbox{M}_\odot$. A full description of the procedure adopted
to generate zoomed initial conditions can be found in Bonafede et al.\ 2011.

In this paper, we will focus only on the two most realistic set of simulations:
CSF, wherein gas cooling, star formation and SN feedback mechanisms are taken
into account (the latter two by means of the sub-grid multi-phase model by
\cite{SH03}), and AGN, in which also the effect of AGN feedback is included.
For a more immediate understanding of the difference between the two sets of
simulations considered here, in the following we will refer to them simply as
AGN OFF and AGN ON respectively.

The identification of clusters has been done by running a FoF algorithm in the
high resolution regions, which links DM particles using a linking length of
0.16 times the mean particle separation. We consider clusters with masses
$M_{200} > 1 \times 10^{14} h^{-1} \mbox{M}_\odot$  at z=0.\footnote{$M_{200}$
($M_{500}$) is the mass enclosed by a sphere whose mean density is 200 (500)
times the critical density at the considered redshift.}.

\subsubsection{Cooling, star formation and stellar feedback}
\label{sec:csf}

Radiative cooling rates are computed following the same procedure presented by
Wiersma et al. (2009). We account  for  the presence of the cosmic microwave
background and of UV/X–ray background radiation from quasars and galaxies, as
computed by Haardt \& Madau (2001). The contributions to cooling from each one
of eleven elements (H, He, C, N, O, Ne, Mg, Si, S, Ca, Fe) have been
pre–computed using the publicly available CLOUDY photo–ionization code (Ferland
et al. 1998) for an optically thin gas in photo–ionization equilibrium. Gas
particles above a threshold density of 0.1 cm$^{-3}$ and below a temperature
threshold of $2.5 \times 10^5$ K are treated as multi-phase, so as to provide a
sub-resolution description of the interstellar medium, according to the model
originally described  by Springel \& Hernquist (2003). The temperature
condition, originally not present, has been introduced here in order to improve
the interaction between the multi-phase star formation model and the AGN
feedback model (see below in this section and Section \ref{app:ene}). Within
each multi-phase gas particle, a cold and a hot-phase coexist in pressure
equilibrium, with the cold phase providing the reservoir of star formation. The
production of heavy elements is described by accounting for the contributions
from SN-II, SN-Ia and low and intermediate mass stars, as described  by
Tornatore et al. (2007). Stars  of different mass, distributed according to a
Chabrier IMF (Chabrier 2003), release metals over the time-scale determined by
their mass-dependent lifetimes. Kinetic feedback contributed by SN-II is
implemented according to the model by Springel \& Hernquist (2003): a
multi-phase star particle is assigned a probability to be uploaded in galactic
outflows, which is proportional to its star formation rate. We assume $v_w =
500$ km s$^{-1}$ for the wind velocity, while assuming a mass-upload rate that
is two times the value of the star formation rate of a given particle.

\begin{figure}
\hspace{-1cm}
 \includegraphics[width=8.5cm, height=8.5cm]{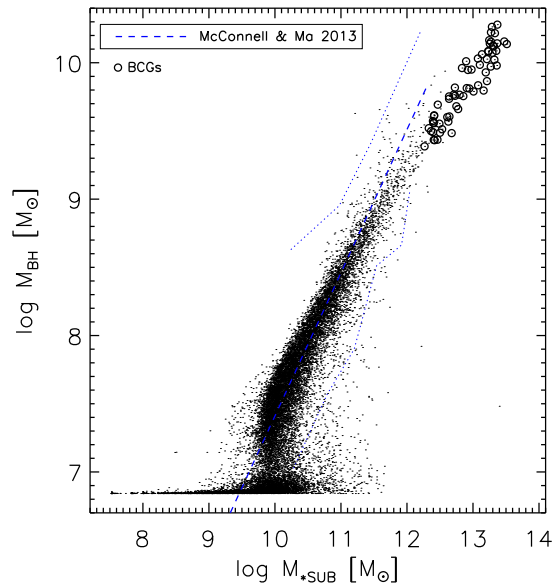}
 \caption{Correlation between the subhalo stellar mass and SMBH mass in our simulations. All subhaloes satisfying the BH seeding condition
 ($M> 2.5 \times 10^{11} h^{-1}$ M$_\odot$)
 have
 been included. The dashed line is a recent
 observational estimate of the $M_{BH}-M_{bulge}$ relation for a sample of 35 galaxies, of mixed morphological type,
 plotted only in the range covered by data (McConnell \& Ma 2013). The two dotted curves
 enclose the region containing the data.
 Open circles refer to BCGs, for which a substantial contribution to $M_{*SUB}$
 arises from intra cluster light. The concentration of points at $\log M_{BH} \sim 6.8$ is due to the mass with which the BH are seeded
 in the subhaloes.
 } \label{fig:calib}
\end{figure}

\subsubsection{AGN feedback}

Our model for the growth of Super-Massive Black Holes (SMBH) and related AGN
feedback is derived from \cite{Springel2005b}, with some differences we
introduced to adapt it to the lower resolution of cosmological simulations. In
this section the model is quickly summarized, while more details, including
justifications for the modifications, can be found in the Appendix.

\begin{itemize}

\item {\it SMBH seeding and growth}. The BHs are represented by means of
    collision-less particles, subject only to gravitational forces. When a
    DM halo is more massive than a given threshold $M_{th}$ and does not
    already contain a SMBH, a new one is seeded with an initial small mass
    of $M_{seed}=5 \cdot 10^6 h^{-1}$ M$_\odot$. We set $M_{th}=2.5 \times
    10^{11} h^{-1}$ M$_\odot$. The SMBH grows with an accretion rate given
    by the minimum between a Bondi accretion rate \citep{Bondi52}, modified
    by the inclusion of a (large) multiplicative factor, and the Eddington
    limit. At variance with respect to the original model we do not
    subtract the corresponding mass from the surrounding gaseous component.
    This results in a small mass non-conservation, which can be neglected
    on galactic scales, while avoiding the gas depletion in the BH
    surroundings, on physical scales much larger (tens of kpc) than its
    physical sphere of influence. This unrealistic removal of gas has,
    among others, also the effect of artificially shallowing the
    gravitational potential in the innermost part of the haloes, making it
    easier for the BHs to drift away from it.

\item {\it SMBH advection and mergers}. In order to avoid numerical
    artifacts, it is fundamental to keep the SMBH at the center of its DM
    halo, counteracting numerical effects that tend to move it away
    \citep{Wurster13}. To obtain this, we reposition at each time-step the
    SMBH particle at the position of the nearby particle, of whatever type,
    having the minimum value of the gravitational potential within the
    gravitational softening of the SMBH. When two BHs are within the
    gravitational softening and their relative velocity is smaller than a
    fraction 0.5 of the sound velocity of the surrounding gas, we merge
    them.

\item {\it Thermal Energy distribution}. In our model, the SMBH growth
    produces an energy determined by a parameter $\epsilon_r$ which gives
    the fraction of accreted mass which is converted in energy. Another
    parameter $\epsilon_f$ defines the fraction of this produced energy
    that is thermally coupled to the surrounding gas. These parameters are
    commonly calibrated in order to reproduce the observed scaling
    relations of SMBH mass in spheroids (see Section \ref{sec:results}).
    Here we set $\epsilon_r=0.2$ and $\epsilon_f=0.2$, which results in a
    reasonable match with a recent estimate of the correlation between
    stellar and BH mass in spheroids (McConnell \& Ma 2013), as shown in
    Fig.\ \ref{fig:calib}. {\rev We explicitly note that these values are
    greater than those adopted by other authors. However, we found that
    decreasing either $\epsilon_r$ or $\epsilon_f$ by a factor $\sim 2$,
    would increase the BH mass by a factor $\sim 2$ for a given stellar
    mass, significantly worsening the match shown in Fig.\
    \ref{fig:calib}}. Note that the simulation points at $\log M_{BH}
    \gtrsim 9.2$ in this figure mostly refer to BCGs (open circles), for
    which a substantial contribution $\sim 50\%$ to $M_{*SUB}$ arises from
    Intra Cluster Light (ICL; e.g.\ Cui et al.\ 2013). A proper estimate of
    the BCG stellar mass (the subject of next Section \ref{map}) yields to
    a better agreement with the data at the high mass end (see Fig.\
    \ref{fig:mstar24_mbh}).

    Following Sijacky et al.\ (2007), we also
    assume a transition from a {\it quasar mode} to a {\it radio mode} AGN
    feedback when the accretion rate becomes smaller than a given limit,
    $\dot{M}_{BH}/\dot{M}_{Edd} = 10^{-2}$. In this case, we increase the
    feedback efficiency $\epsilon_f$ by a factor four.

    In the original model of \cite{Springel2005b}, the resulting energy was
    simply added to the specific internal energy of gas particles. However,
    due to the features of the adopted star formation and stellar feedback
    multi-phase model \citep{SH03}, when this energy is given to a star
    forming particle, it is almost completely lost without practical
    effects on the system (see Appendix \ref{app:ene}). To avoid this,
    whenever a star-forming gas particle receives energy from a SMBH, we
    calculate the temperature at which the cold gas phase would be heated
    by it\footnote{The AGN energy is given to the hot and cold phases
    proportionally to their mass}. If this temperature turns out to be
    larger than the average temperature of the gas particle (before
    receiving AGN energy), we consider the particle not to be multi-phase
    anymore and prevent it from forming stars. To avoid an immediate
    re-entering of this gas particle in multi-phase star forming state, we
    add to the usual minimum density condition, a maximum temperature
    condition for a particle to be star forming, as anticipated in Section
    \ref{sec:csf}.
\end{itemize}

\begin{figure*}
\includegraphics[width=7.8cm, height=8.2cm]{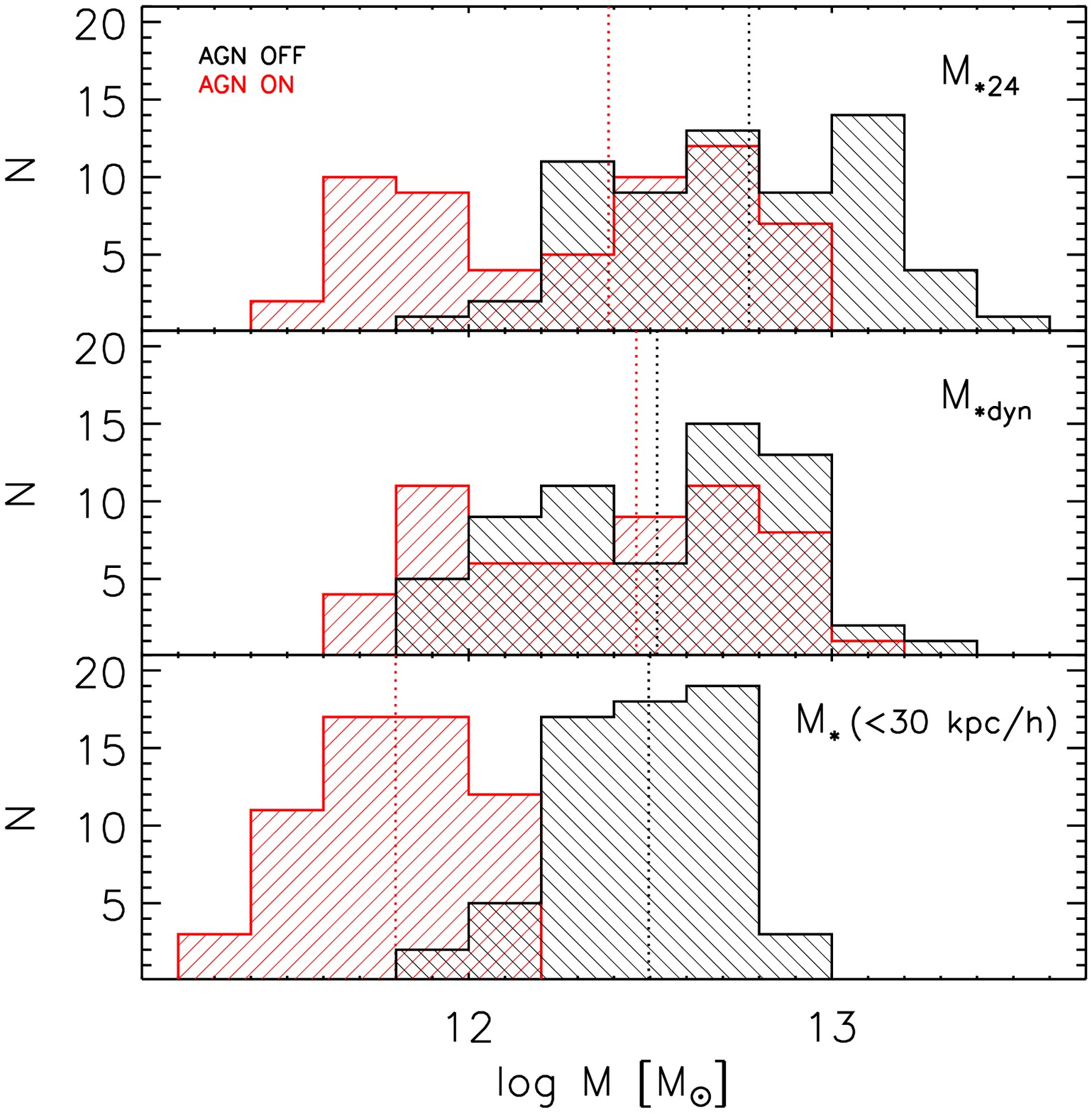}
\hspace{0cm}
\includegraphics[width=8.5cm, height=8.5cm]{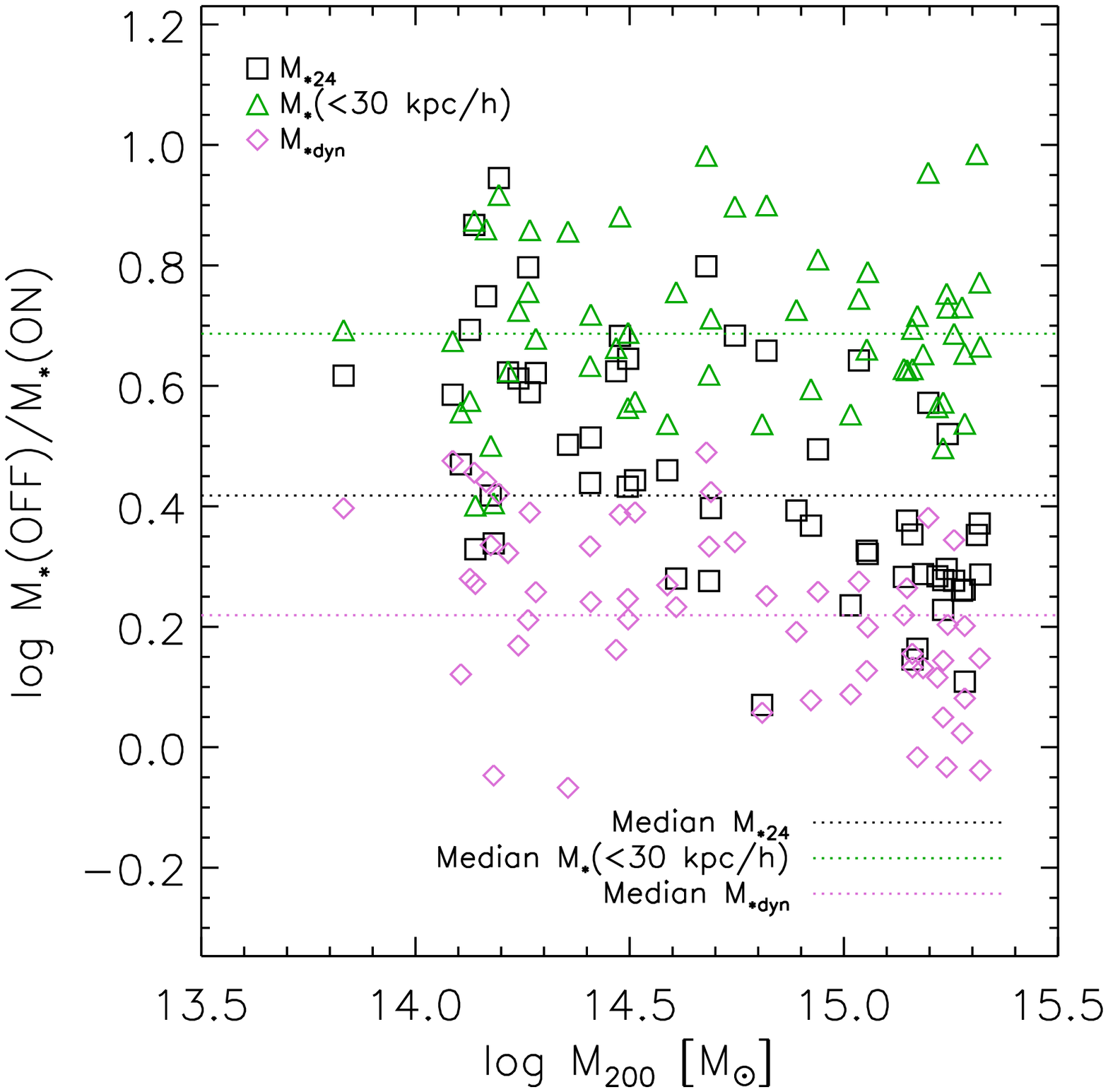}
 \caption{Left panel: histograms of the stellar masses of the BCGs, according to three different
 definitions
 (see text) for runs with AGN ON and OFF.
 The vertical lines show the median of the distributions.
 Right panel: ratios between the masses of the simulated BCGs without and with AGN feedback
 included, as a function of halo mass, for the same three definitions of BCG mass. The horizontal lines mark the
median, with the same colours as the points
  for the three definitions.
 } \label{fig:masas}
\end{figure*}

\subsection{Surface brightness maps}
\label{map}

To generate mock maps of the clusters, we follow the same procedure as in Cui
et al.\ (2013) and Cui et al.\ (2011). Each star particle is treated as a
Simple Stellar Population (SSP) with age, metallicity and mass given by the
corresponding particle properties in the simulation, and adopting the same
initial mass function (IMF) (a Chabrier IMF; Chabrier 2003). The spectral
energy distribution (SED) of this particle is computed by interpolating the SSP
templates of Bruzual \& Charlot (2003), and a standard Johnson V-band filter is
applied to this SED, to get its V-band luminosity. Then, we smooth to a 3D mesh
grid both the luminosity and the mass of each star particle with a SPH kernel.
The mesh size is fixed to $5 \Kpc$ (corresponding to the simulation's softening
length). Finally, by projecting this 3D mesh in one direction we obtain the
mock 2D photometric image. The procedure neglects dust reprocessing.

The simulated maps of the clusters have been used to compute isophotal radii,
masses within these radii and half light (effective) radii of BCGs. In
particular, we are interested in defining the total extent and mass of galaxies
from mock maps, emulating what would be done using real observations. This
is notoriously a difficult, if not an ill-posed, problem. Here, we generally
follow a commonly adopted definition of the outer limit of galaxies, namely the
isophote 25 mag arcsec$^{-2}$ in the B band, which for BCGs translates to about
24 mag arcsec$^{-2}$ in the V band. This isophotal boundary is at the basis of
the D25 diameter and corresponding R25 radius, widely used in literature since
its introduction by de Vaucouleurs et al.\ (1976). However, it is well known
that this limit tends to underestimate the real extension of galaxies, so that
we have also checked that our conclusions are substantially unchanged, and if
anything strengthened, by adopting flux limits $\sim 1$ mag fainter.

\begin{figure*}
\includegraphics[width=9.0cm, height=8.5cm]{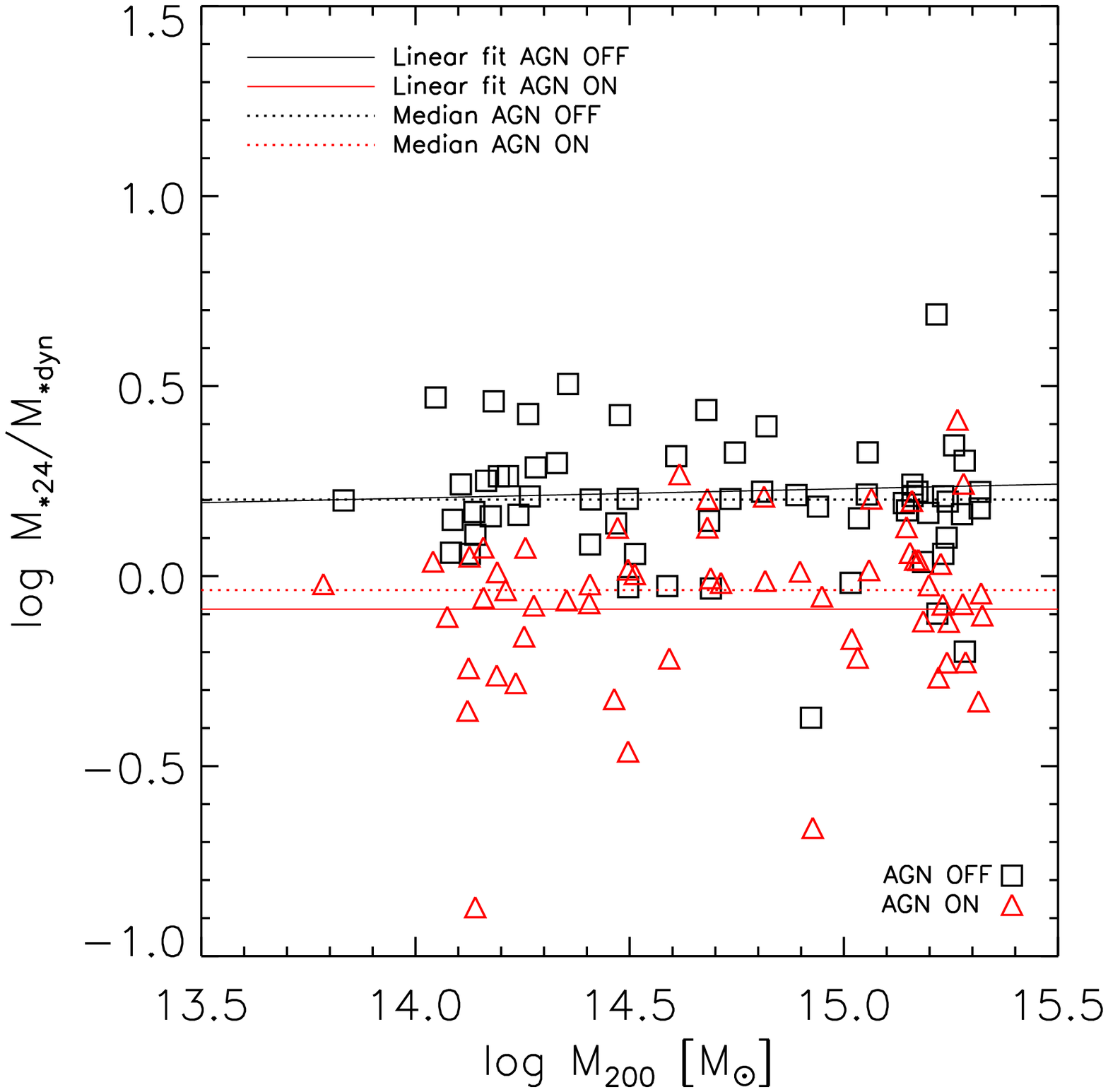}
\hspace{-1cm}
\includegraphics[width=9.0cm, height=8.5cm]{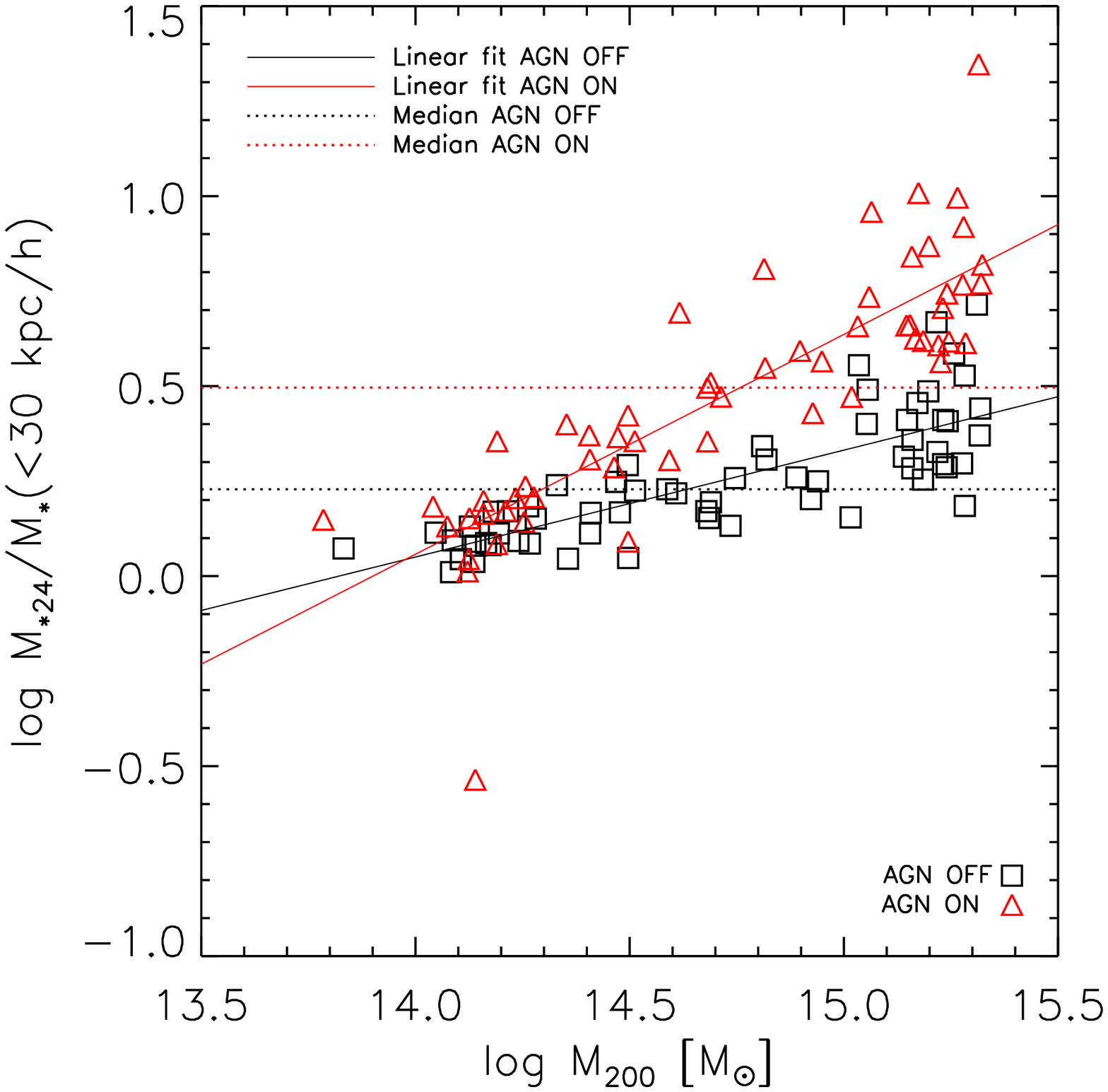}
  \caption{Left panel: Ratios between the BCG mass enclosed within the 24 mag arcsec$^{-2}$ in the V band, and that
   obtained by means of a dynamical separation from ICL (Cui et al.\ 2013;
  see text for more details) as a function of the halo mass. Right
  panel: the same but with the former mass replaced by the stellar mass within a radius
  of 30 kpc/$h$.
 } \label{fig:mass_ratios_24dyn30}
\end{figure*}

\section{Results}
\label{sec:results}

\subsection{The effect of AGN feedback on BCG mass}

For almost a decade, AGN feedback has begun to be taken into account in galaxy formation,
in order to avoid the prediction of too massive and too young galaxies.
Actually, its relevance is quite plausible from
a physical point of view. For instance, it is easy to see that, given the
observed ratio of SMBH to stellar mass in spheroids, it is sufficient a very
small fraction (a few percent), of the energy produced by accretion onto the
SMBH to totally unbind the gas from the potential well of the galaxy. On the
other hand, it has been well assessed that computations without some
efficient inhibition of stellar mass assembly at the high mass end of the
luminosity function,  where SNae feedback becomes increasingly insufficient,
over-predict the stellar mass by a factor $\gtrsim 10$ (e.g.\ Benson et al.\
2003). Negative feedback coming from AGN activity helps in reducing, if not
totally solving, this problem.

In the left panel of Fig.\ \ref{fig:masas} we show the distributions of the
stellar masses of the BCGs, for runs with AGN ON and OFF, while the right panel
displays the ratio between the masses of the same BCGs in AGN OFF and AGN ON
simulation, as a function of hosting halo mass. Here we adopt three different
possible definitions of the stellar mass. $M_{*24}$ is the mass enclosed by the
24 V mag/arcsec$^2$ isophote, mimicking a commonly adopted procedure to
estimate total galaxy masses from observed images, as discussed in Section
\ref{map}. $M_{*dyn}$ is instead the BCG mass computed performing a dynamical
separation from the ICL, or more precisely from star particles not
gravitationally bound to the galaxy, by means of an analysis of the velocity
distribution. In brief, the method consists in recognizing that the latter is
best fitted by the superposition of two Maxwellians, and in ascribing the low
and high velocity components to the BCG and ICL respectively. Individual
stellar particles are then assigned to the two components by means of an
analysis of their velocities (see Cui et al.\ 2013 for details and references).
Finally, $M_*(<30 \ \mbox{kpc}/h)$ is the mass enclosed within a fixed radius
from the galaxy center (defined as the center of mass of the stellar
component). This simple choice of estimating the mass within a fixed physical
radius has often been adopted to extract BCGs in hydro simulations. For
instance McCarthy et al.\ (2010) and Stott et al.\
(2012)\footnote{\label{note_stott} Note that while in Stott et al.\ (2012) the
masses are stated to be computed within 30 kpc, they were actually computed
within 30 kpc/h (Stott, private comunication).} adopted 30 kpc$/h$, while
Sijacki et al.\ (2007) used 20 kpc/$h$. However, this criterion is not well
justified from an observational point of view (see below).

The two panels of  Fig.\ \ref{fig:masas} show that the inclusion of the AGN
feedback indeed decreases the stellar mass. However the average decrease is
only by a modest factor $\sim$ 1.5-5, depending on the adopted definition of
mass. As we will see in more detail below, this is helpful but still
un-sufficient for a satisfactory comparison with real galaxies. For $M_{*24}$,
i.e.\ the mass obtained by means of a procedure close to an observational one,
(and to a lesser extent also for $M_{*dyn}$) the decrease tends to be less
pronounced in more massive haloes.



Fig.\ \ref{fig:mass_ratios_24dyn30}, left panel (see also left panel of Fig.\
\ref{fig:masas}), illustrates that $M_{*24}$ is on average $\sim 60$\% larger
than $M_{*dyn}$ for AGN OFF simulations, but only $\sim 15$\% smaller for AGN
ON runs, and that the ratios between these two masses do not have a clear trend
with the halo mass. Indeed, Cui et al.\ (2013), analyzing this same set of
simulations, find that the brightness profile of stars not dynamically bound to
the BCG (i.e.\ not entering in the computations of $M_{*dyn}$), begins to
dominate over that of the bound component on average at $\mu_V\gtrsim 23$
mag/arcsec$^2$ for AGN OFF runs, with a weak dependence on mass. This limit
turns out to be brighter enough than our adopted isophotal boundary
($\mu_V\gtrsim 24$ mag/arcsec$^2$) to exclude a fraction $\sim 50\%$ of the
mass included by the latter. Conversely, the transition occurs at
($\mu_V\gtrsim 24.75$) for AGN ON runs, almost independent of mass, a value
fainter than our limit, which however does not add much mass.

Also the ratios between $M_{*24}$ and $M_*(<30\ \mbox{kpc}/h)$ depend on the
flavor of the simulation, being on average over the whole sample $\sim 1.7$ and
$\sim 3$  for AGN OFF and for AGN ON respectively (see right panel of Fig.\
\ref{fig:mass_ratios_24dyn30} and also left panel of Fig.\ \ref{fig:masas}).
However these ratios feature a quite evident trend with the halo mass, ranging
from slightly more than 1, up to 2 and 5 respectively. Therefore, $M_*(<30\
\mbox{kpc})$ in general underestimates the BCG stellar mass, both with respect
to the surface brightness limit and to the dynamical criterion. The problem
worsen for AGN ON runs, since AGN feedback, besides making galaxies less
massive, expands them (see Fig.\ \ref{fig:mstar_re} and related discussion
below). Thus the use of a fixed physical radius to estimate stellar quantities
introduces a bias, depending on the mass and on the sub-grid physics included.
In particular, for the scope of the present paper, it can overstate the
effectiveness of the AGN feedback in limiting the mass growth of galaxies,
especially at high masses. We will come back to this point in Section
\ref{subsec:halo_sf_eff}.

For the above reasons,  when comparing our results with data on total masses of
BCGs, we adopt in the following $M_{*24}$, which is obtained mimicking a
procedure often adopted in analyzing observations. However, it is reassuring to
note that our conclusions would not be substantially affected by the use of
$M_{*dyn}$.


\begin{figure}
  \centerline{\includegraphics[width=9.5cm, height=8.5cm]{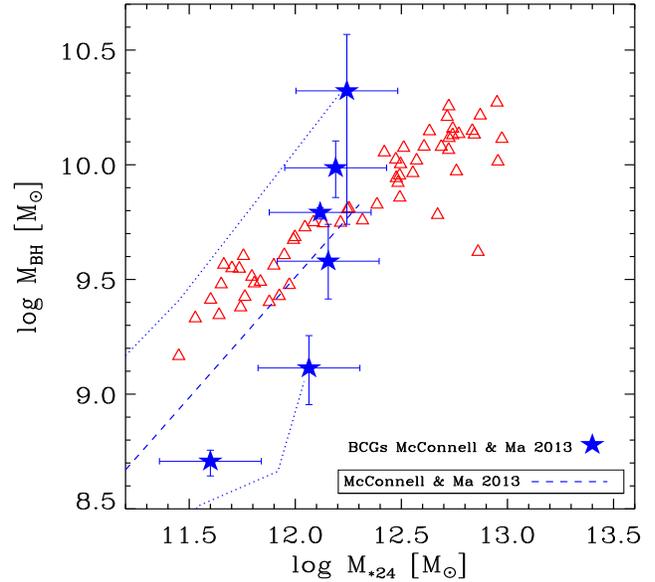}}
  \caption{Correlation between the stellar and SMBH masses compared to a recent
  observational determination by McConnnell and Ma (2013),
  for a mixed sample of 35 galaxies with dynamically estimated bulge masses, and including 6 BCGs at the high mass end,
  plotted only in the range covered by data. The two dotted curves
 enclose the region containing all the available data, of which we plot only BCGs (blue stars).
 } \label{fig:mstar24_mbh}
\end{figure}

\begin{figure}
  \centerline{\includegraphics[width=9.5cm, height=8.5cm]{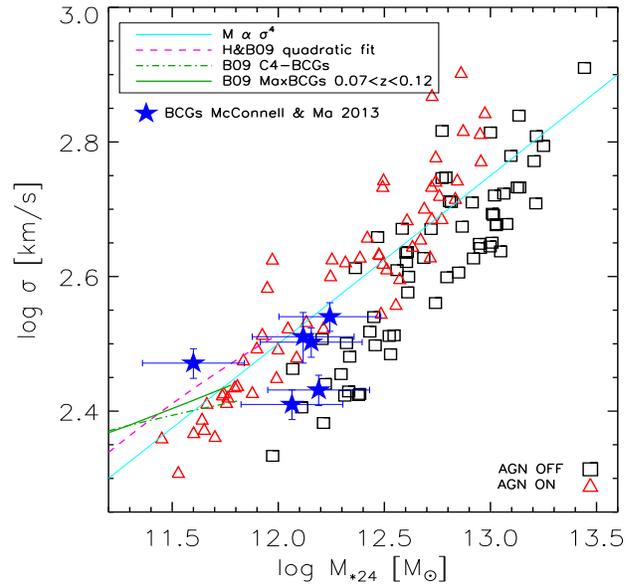}}
  \caption{Correlation between the stellar mass and central (within $R_e/8$) 1-d
  velocity dispersion for our mock BCGs (Faber-Jackson relation), compared to recent estimates by
  Hyde \& Bernardi (2009; H\&B09) for elliptical galaxies and by Bernardi (2009; B09) for BCGs.
  These are plotted only in the range covered by their BCGs. The runs with and without AGN feedback are in
  red and black respectively.
}
  \label{fig:faber}
\end{figure}

\begin{figure}
  \centerline{\includegraphics[width=9.5cm, height=8.5cm]{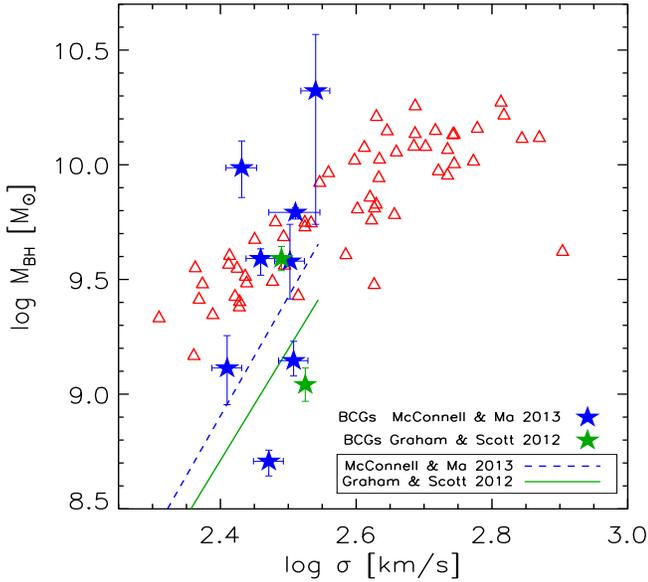}}
  \caption{Correlation between the SMBH masses and stellar central (within $R_e/8$) 1-d
  velocity dispersion for our mock BCGs, compared to the determination by McConell \& Ma (2013)
  for 53 Early Type Galaxies, including 8 BCGs, and that by Graham \& Scott (2013) for 28 Ellipticals, two
  of which are BCGs.}
  \label{fig:sigma_mbh}
\end{figure}

\subsection{SMBH mass vs BCG mass and velocity dispersion}

When implementing in simulations sub resolution prescriptions for the growth of
SMBH and for the ensuing feedback, the first basic requirement is to reproduce
the observed correlation between the SMBH mass and the velocity dispersion, or
the stellar mass, found for the spheroidal component of galaxies. Actually, the
parameters entering in the sub-grid prescription for SMBH growth and AGN
feedback, namely the accretion efficiency, the coefficient $\alpha$ in front of
the Bondi accretion rate and the fraction of accretion energy thermally
released to the interstellar medium (Eqs.\ \ref{eq:bondi}, \ref{eq:edd} and
\ref{eq:feedback}), have been calibrated here, as well as in previous works, in
order to reproduce the $M_{BH}-M_{*}$ relationship, over the whole mass range
accessible with the simulation (see Figure \ref{fig:calib} and Section
\ref{simulation}).

In view of this,  Fig.\ \ref{fig:mstar24_mbh} unsurprisingly shows that, once
the stellar mass of the BCGs has been properly computed, correcting $M_{*SUB}$
for the ICL contamination, their relationship between the stellar and SMBH mass
turns out to be in better agreement with that derived from observations,
although with a somewhat shallower slope ($M_{BH} \propto M_{*}^\alpha$ with
$\alpha \sim 0.7$ while $\alpha \sim 1$ in observation). Here we compare in
particular our simulated BCGs with the recent observational determination by
McConnnell \& Ma (2013), for a mixed sample of 35 galaxies with dynamically
estimated bulge masses, and including 6 BCGs, at the high mass end.
Unfortunately, these BCGs have all (but one) similar masses within a factor
$\lesssim 2$. As a consequence, while there is mounting evidence that these
special objects differ from the general early type galaxy population for some
other scaling relations (e.g.\ Hyde \& Bernardi 2009; Berardi 2009), no
conclusion can be drawn for the $M_{BH}-M_*$ relation. The simulated points
extend to masses larger than observed by a factor $\sim 4$. This is a first
general indication of the fact that even in our AGN ON runs the formation of
baryonic structures is not prevented enough. This point will be addressed in
Section \ref{subsec:halo_sf_eff}. Note that, although we find in simulations a
shallower slope than that found considering the whole early type galaxy
population, the few observed BCGs scatter quite evenly around the region
defined by the simulated points. Also, the intrinsic dispersion of real
galaxies around the relationship is larger than that of simulated galaxies, as
can be appreciated also from Figure \ref{fig:calib}. This can be due to an
underestimate of the observational uncertainties, or to a physical link between
star formation and SMBH accretion somewhat looser than that assumed in our
simulation.


Since the '70s, it has been known that the luminosity, or stellar mass, of
early type galaxies strongly correlates with their velocity dispersion, $M_*
\propto \sigma^\gamma$ with $\gamma \simeq 4$ (Faber \& Jackson, 1976). More
recently, it has been pointed out that this relationship shows a {\it
curvature} at the high mass end, and in particular for BCGs, in the sense that
the increase of $\sigma$ with stellar mass becomes slower, with $\gamma \simeq
5-6$ (Hyde \& Bernardi 2009; Bernardi 2009). Fig.\ \ref{fig:faber} shows the
simulated Faber-Jackson relationship for AGN ON and AGN OFF runs, where
$\sigma$ is the one dimensional velocity dispersion computed within $R_e/8$.
The slopes are similar but in the former case the normalization is closer to
observations. In both cases we get an increase of velocity dispersion with
stellar mass described by $\gamma \sim 3$, faster than that of the general
population of early type galaxies. This disagreement becomes more important if
we consider the study by Bernardi (2009), based on two samples of thousands of
BCGs. Moreover, there are no known BCGs with $\sigma > 400$ km/s (e.g.\ von der
Linden et al.\ 2007), while several simulated BCGs have $\sigma$ well above
this limit. An important point to notice is that our AGN feedback, while
decreasing the final stellar mass by a factor of a few, affects very little the
predicted stellar velocity dispersion. In other words, the inclusion of AGN
feedback moves the points in Fig.\ \ref{fig:faber} almost horizontally to the
left. Thus, while one could possibly imagine to reduce the residual over
prediction of stellar masses with an even more aggressive choice of parameters,
it is unlikely that this would help in making the velocity dispersions closer
to the observed values. We will come back to this issue in Section
\ref{sec:discussion}.


The correlation between the SMBH mass and the central velocity dispersion
$\sigma$ is shown in Fig.\ \ref{fig:sigma_mbh}, and compared to recent
observational determinations. McConnell \& Ma (2013) provide the fit for a
sample of 53 early type galaxies, including 8 BCGs, while Graham \& Scott
(2013) give a bisector regression for a sample of 28 ellipticals, two of which
are BCGs. In both cases, the BCGs occupy a narrow $\sigma$ range, at the high
end. These recent observational estimates point to a scaling of SMBH mass with
$\sigma$ as steep as $M_{BH} \propto \sigma^5$, while our simulated BCGs are
clearly characterized by a much shallower power law slope $\sim 2$. This is in
keeping with the findings, presented in Fig.\ \ref{fig:mstar24_mbh} and Fig.\
\ref{fig:faber}, that in our BCGs $M_* \propto \sigma^3$ and $M_{BH} \propto
M_{*}^{0.7}$.  As a result, while in the range of $\sigma$ covered by
observations the simulated SMBH masses are larger than observed, if the
observed correlation is extrapolated to larger masses, the opposite becomes
true. Although the simulated point follow a quite shallower $M_{BH}-\sigma$
relationship than that defined by the whole early type galaxy population, the
few observed BCGs are distributed around the former with a larger dispersion,
similarly to the case of the $M_{BH}-M_*$

\begin{figure}
  \centerline{\includegraphics[width=9.5cm, height=8.5cm]{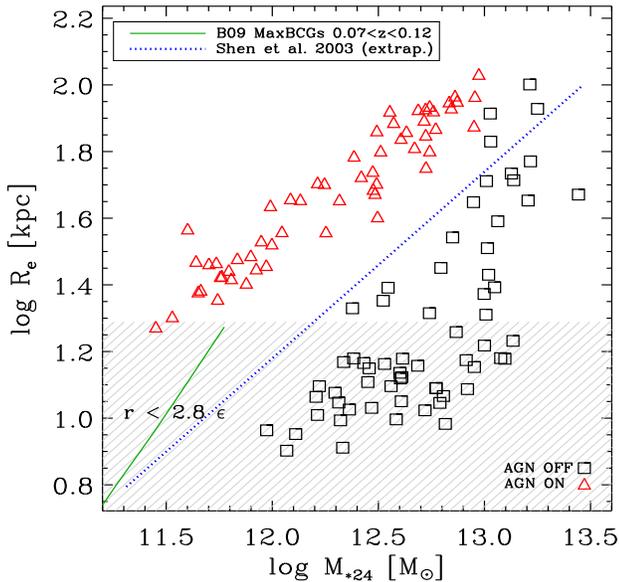}}
\caption{Effective radius against total stellar mass,
computed within 24 magnitude in the V band, compared to the observed correlation
in BCG (Bernardi 2009; plotted only in the region constrained by data), and to
the extrapolation of the correlation derived for the global early type galaxies
population by Shen et al.\ (2003). The runs with and without AGN feedback are in red and black respectively.
The hatched area highlights the scale $2.8 \epsilon$ over which, due to softening,
the gravitational force deviates from the pure Newton law, although
in most papers $\epsilon$ is regarded as the spatial resolution limit.
 } \label{fig:mstar_re}
\end{figure}


\subsection{Mass-size relation}

Normal Early Type Galaxies (ETGs) obey a well defined relationship between
their stellar mass and effective radius $R \propto M^{0.6}$ (Mass-Radius
Relation or MRR), as pointed out already by Burstein et al.\ (1997), and later
substantially confirmed by various authors on the basis of the Sloan Digital
Sky Survey (SDSS). As for BCGs, there have been both claims for a steeper slope
$R \propto M^{0.9}$ (e.g.\ Bernardi 2009, Guo et al.\ 2009), as well as for no
substantial deviation with respect to normal ETGs (von der Linden et al.\
2007). In Fig.\ \ref{fig:mstar_re} we plot the effective radius against total
stellar mass. The latter has been computed within the 24 mag/arcsec$^2$ V band
isophote, for our mock BCG images, following a common observational definition
(Section \ref{map}). In the same vein, the effective radius has been evaluated
as the mean radius of the isophote enclosing half of the total light within the
24 mag/arcsec$^2$ V. The {\rev blue dotted line line} is an extrapolation (at
masses $> 10^{12} M_\odot$) of the MRR relationship for ETGs by Shen et al.
2003, the {\rev short green line} on the left is the fit given by Bernardi
(2009) for BCGs in the redshifts bin $0.07<z<0.12$. The latter is plotted only
in the range of masses in which it is actually derived, meaning that there are
no BCGs in her sample whose stellar mass is larger than $10^{12}$ M$_\odot$.
The figure shows that SMBHs feedback, besides making the galaxies less massive,
with a greater effect at low masses, induces also a significative expansion by
an average factor $\sim 4$. As a result, while the AGN OFF points lies below
the extrapolations of both observed MRRs, the AGN ON are mostly above the Shen
et al.\ (2003) MRR. By converse, the steeper law found by Bernardi (2009) for
BCGs, when extrapolated to our mass range, would predict even larger galaxies.
It is also interesting to note that the dispersion of the simulated correlation
is much smaller in the AGN ON case. This supports, from a different point of
view, the suggestion made by Ragone-Figueroa \& Granato (2011).  Using
controlled numerical experiments, they proposed that AGN driven winds could
have a role in explaining the low scatter of the observed local mass-size
relationship (Shen et al.\ 2003; Bernardi 2009). This contribution may arise
from the trade-off they found between variations of the assumed pre-wind galaxy
size, and the amplification due to the wind driven mass loss. As a result, the
post-wind size was found to be relatively insensitive to the initial one (see
their figure 6).

Note that while AGN feedback clearly increases the effective radius, the same
does not happen to the size of the 24 mag/arcsec$^2$ V band isophote, which we
use to define the galaxy boundary. Indeed, we will see in Section
\ref{subsec:prof} that the stellar density profiles are significantly flatter,
with hints for a core, in the AGN ON runs. Our AGN feedback greatly affects the
stellar density in the central regions, but little in the galaxy outskirts.

\begin{figure*}
\includegraphics[width=9.0cm, height=8.5cm]{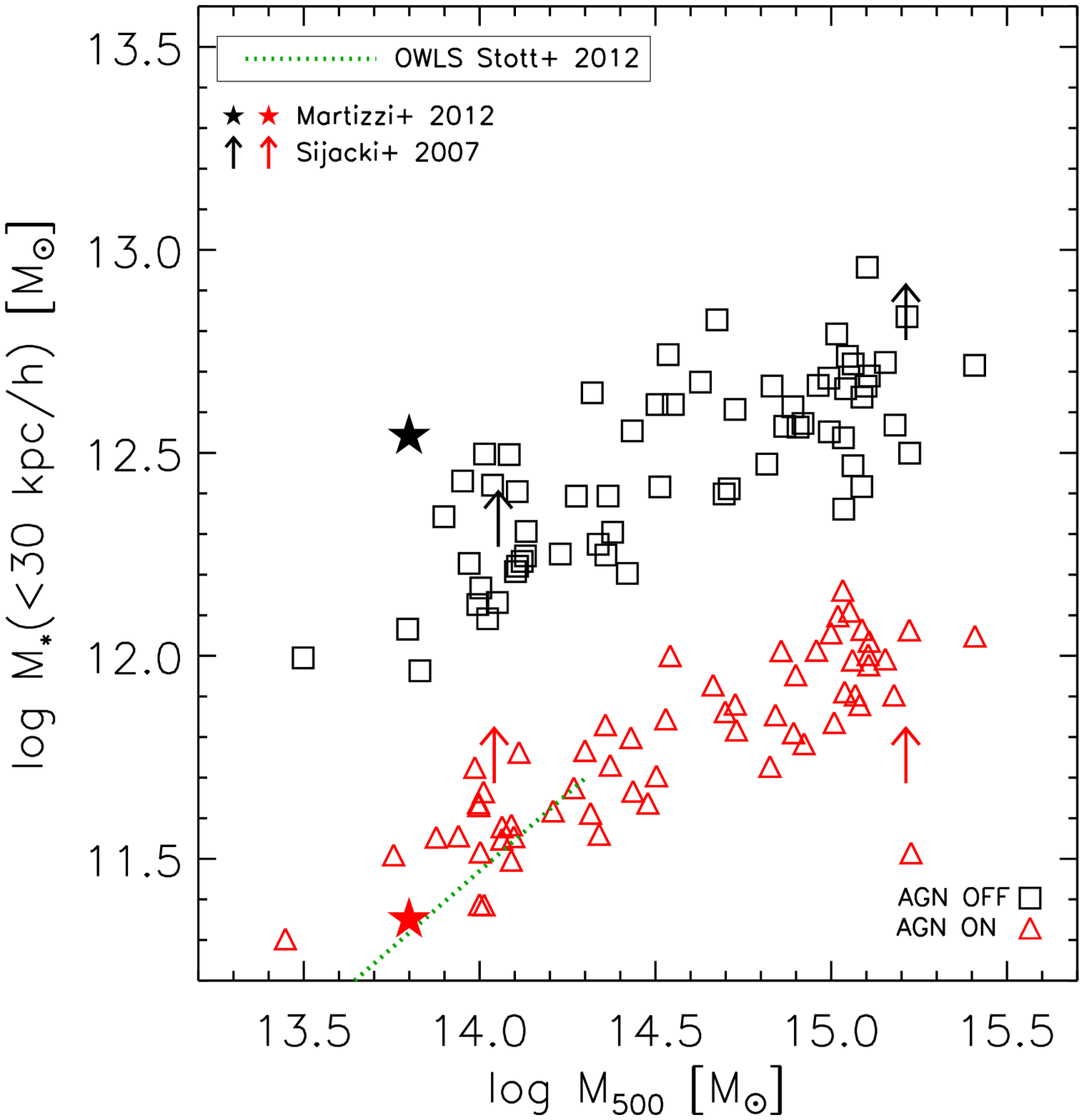}
\hspace{-1cm}
\includegraphics[width=9.0cm, height=8.5cm]{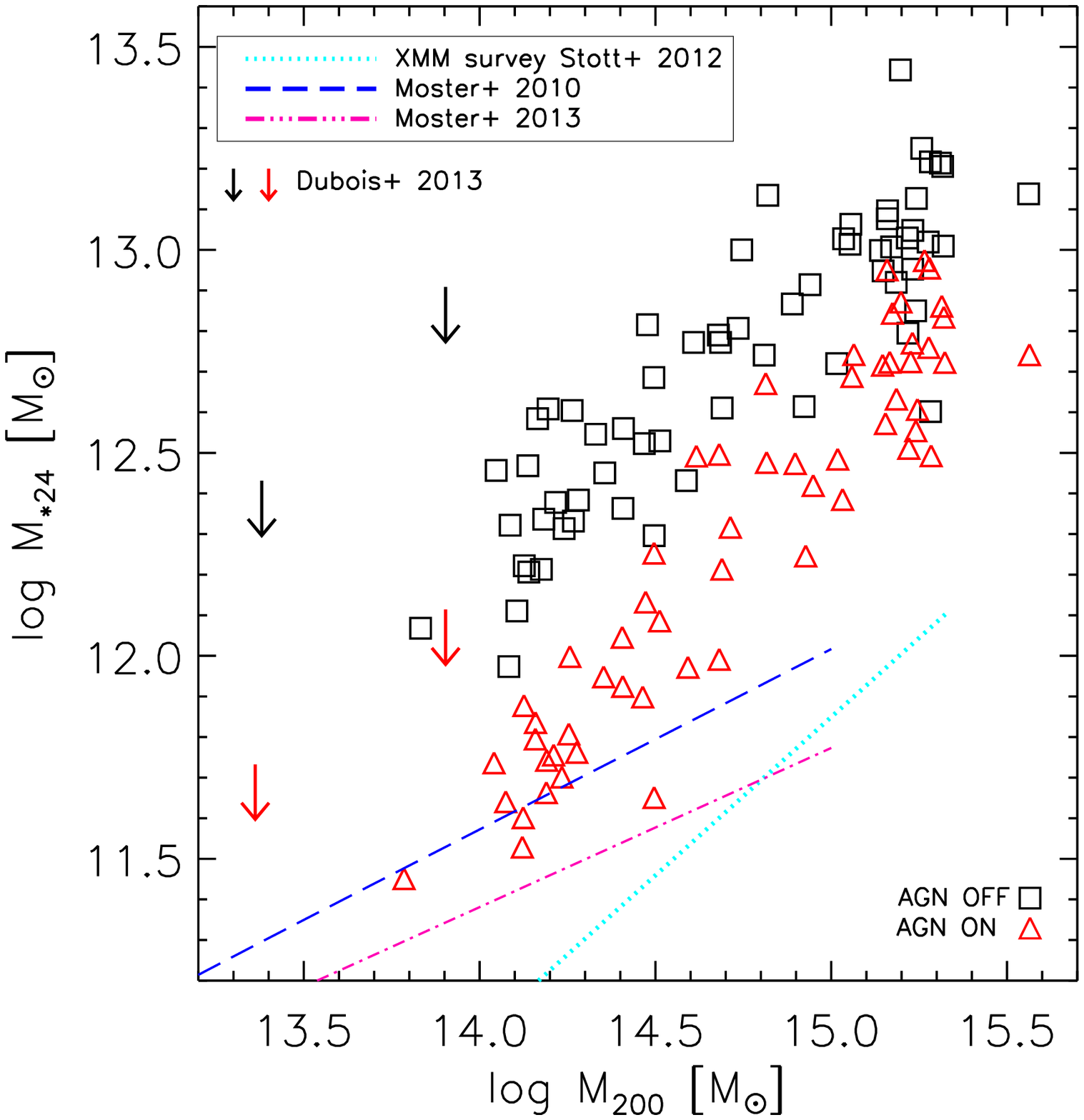}
  \caption{Left panel: stellar mass within 30 kpc/$h$ (as in Stott et al.\ 2012 for
  the OWLS simulation, green line) against $M_{500}$. Upward arrows mark
  the mass within 20 kpc/$h$ (hence lower limits in this plane, in particular for the high mass object)
  from the cosmological simulations of two galaxy clusters by Sijacki et
  al. 2007. Stars show the results by Martizzi et al. 2012a. They identified  the BCG
  by means of a minimum stellar density criterium, but for
  their relatively small mass object the corresponding mass do not differ from $M_{*}(< 30 \
  \mbox{kpc}/h)$ by more than 10\%.
  Right panel: Stellar mass $M_{*24}$ enclosed by the 24 V
  mag/arcsec$^2$ isophote (a commonly adopted boundary for galaxies in observations) vs $M_{200}$, compared with
  recent abundance matching determinations (blue and purple lines) and with results from the XMM cluster survey (cyan line).
  All lines in these plot are plotted only in the region constrained by the
  data. The arrows show results of simulations by Dubois et al.\ 2013,
  for their two most massive galaxy groups. These are to be regarded as upper limits in this plot, since their estimate of the
  stellar mass of the central galaxy includes also the intra-cluster light. For all simulations, the runs with and without AGN feedback
  are plotted in red and black respectively.
 } \label{fig:m500_mstarmin30}
\end{figure*}

\subsection{The halo star formation efficiency}
\label{subsec:halo_sf_eff}

In Fig.\ \ref{fig:m500_mstarmin30}, left panel, we compare our results on the
relationship between the BCG stellar mass within 30 kpc/h and the halo mass
$M_{500}$, with those by Stott et al.\ (2012) for the OWLS simulation\footnote
{see footnote \ref{note_stott}}. The latter simulations are also run with
GADGET-3, but including a a different scheme for AGN feedback (Booth \& Schaye
2009). In the mass range $ 3 \times 10^{13} < M_{500}/\mbox{M}_{\odot} <2
\times 10^{14} $, in which there is superposition between the two simulation
sets, the agreement is good. In the higher mass regime covered by our
simulations,  there is a clear indication for a shallowing of the slope.

{\rev In the same plot, we have also reported the results obtained with the
cosmological simulations of two galaxy clusters by Sijacki et al. 2007. We mark
them as lower limits in this plane, since they estimated the BCG mass as the
mass within 20 kpc/$h$. Finally, we include the results (stars) by Martizzi et
al. 2012a, who identified the BCG by means of a minimum stellar density
criterium. They claim that the corresponding mass does not differ from $M_{*}(<
30 \ \mbox{kpc}/h)$ by more than 10\%.}

We have already remarked (see discussion of Fig.\
\ref{fig:mass_ratios_24dyn30}) that the stellar mass computed within 30 kpc/$h$
is in general an underestimate of simulated BCGs total stellar mass, and that
this underestimate is more severe for more massive galaxies, and when AGN
feedback is included. As a consequence, its use provides an optimistic measure
of the effectiveness of this physical process in limiting the stellar mass. As
for observations, it is worth mentioning that Stott et al.\ (2011) found, for a
local sample of BGCs, a mean value of the effective radius of $43.2$ kpc.
Moreover, for a typical early type galaxy profile, the half mass radius is even
larger than the half light radius. Thus the stellar mass within 30/$h\sim40$
kpc in real BCGs is on average significantly less than half of the total mass.
In conclusion, the practise of computing BCG stellar masses from simulations
within $\sim 30-40$ kpc, in order to compare with total BCGs masses derived
from observations is misleading.

For these reasons, in the right panel of Fig.\ \ref{fig:m500_mstarmin30} we
compare with observations $M_{*24}$ against $M_{200}$. {\rev The arrows show
results of simulations by Dubois et al.\ 2013, for their two most massive
galaxy groups. They are reported as upper limits, since their estimate of the
stellar mass of the central galaxy includes also the intra-cluster light.} The
cyan dotted line is a fit to observational estimates of the total stellar mass
(from profile fitting) for the XMM cluster survey, given by Stott et al.\
(2012). We also plot recent determinations of this relationship based on the
abundance matching technique (Moster et al.\ 2010, 2013). These are
characterized by a slope, in the relevant mass range, much shallower than that
found by Stott et al.\ (2012), and more in keeping with some other
observational determinations (e.g. Lin \& Mohr 2004; Popesso et al.\ 2007; Guo
et al.\ 2009). If this slope is confirmed, the figure shows that our
simulations, even with AGN feedback included, increasingly over-predict the BCG
stellar mass with increasing halo mass. The use of $M(<30 \ \mbox{kpc}/h)$ to
estimate the BCG mass would largely mask the problem. In the low mass region,
AGN ON simulations produce an amount of stars in relative agreement with
observational estimates.

{\rev Our finding that AGN feedback helps to decrease the masses of BCGs, but
falls short of reducing them to the observed values in massive clusters, is in
good agreement also with the results shown by Puchwein et al.\ 2010 for BCGs
luminosities (see their figure 8), and by Puchwein \& Springel 2010 (their
figure 5). To exacerbate the problem, we note that the over-prediction of
stellar mass in our, and most other simulations, could be in part artificially
reduced by the use of SPH technique to treat hydrodynamics. Indeed, it has been
claimed that this method suffers a form of "numerical quenching" of cooling,
particularly important in large haloes. Indeed, Vogelsberger et al.\ (2013)
found that, using the novel moving mesh method for the hydrodynamical forces,
it is necessary to invoke {\it very} strong forms of stellar and AGN feedback
in order to reasonably reproduce the local galaxy luminosity function in
cosmological simulations. More specifically, they had to increase significantly
the strength of radio-mode feedback, compared to previous studies, and to adopt
an energy per SNII event three times larger than the canonical one. Still, the
bright end of the luminosity function remains somewhat overproduced.}

\begin{figure*}
  \centerline{\includegraphics[width=16cm, height=14cm]{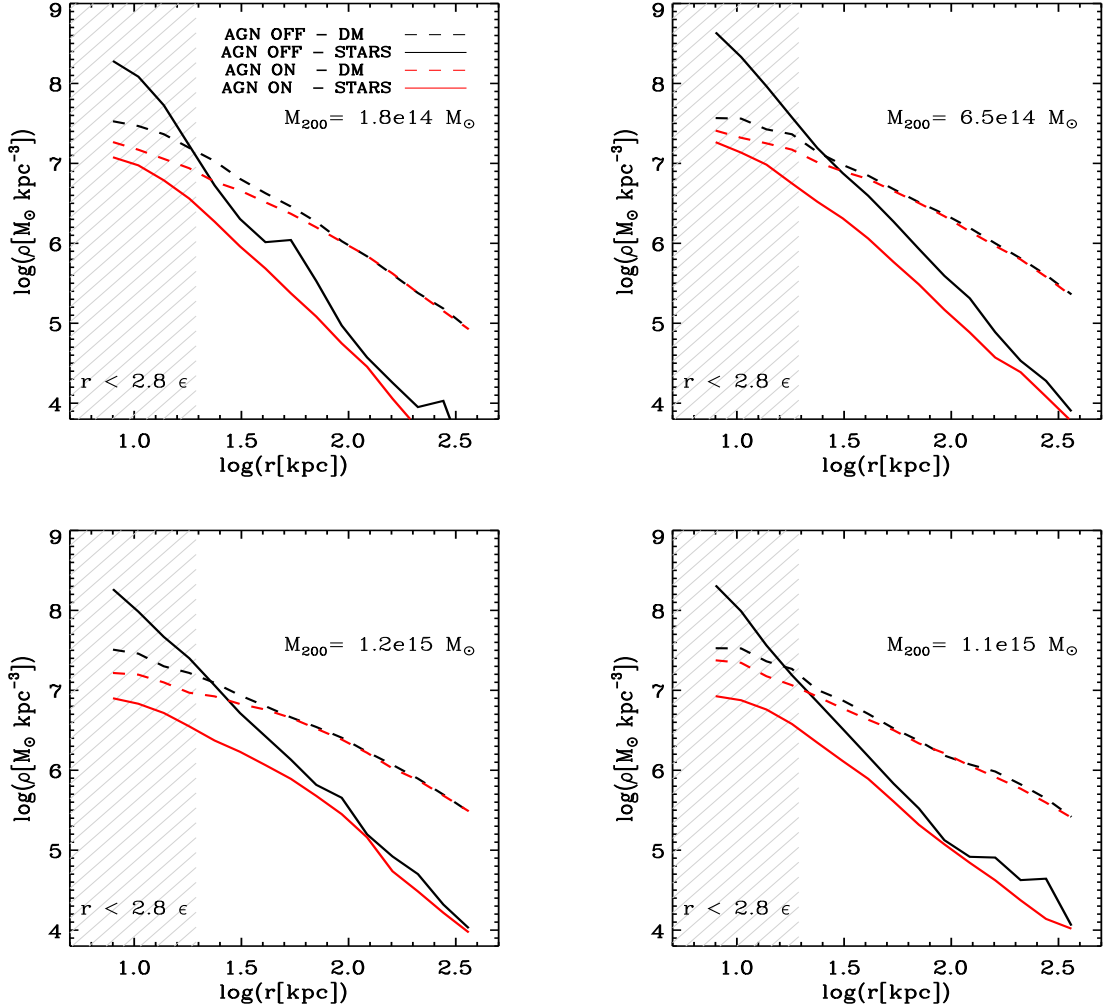}}
  \caption{Density profiles for 4 representative clusters. Is is
  apparent that AGN feedback produces a significant flattening
  of the inner stellar and, to a lesser extent, DM  profiles. The
  hatched areas highlight the scale $2.8 \epsilon$ over which, due to softening,
the gravitational force deviates from the pure Newton law. Note that in most
papers $\epsilon$  (where the plotted lines begin) is regarded as the spatial
resolution limit.
   }
\label{fig:perf_bhnew}
\end{figure*}

\begin{figure}
  \centerline{\includegraphics[width=9cm, height=8.5cm]{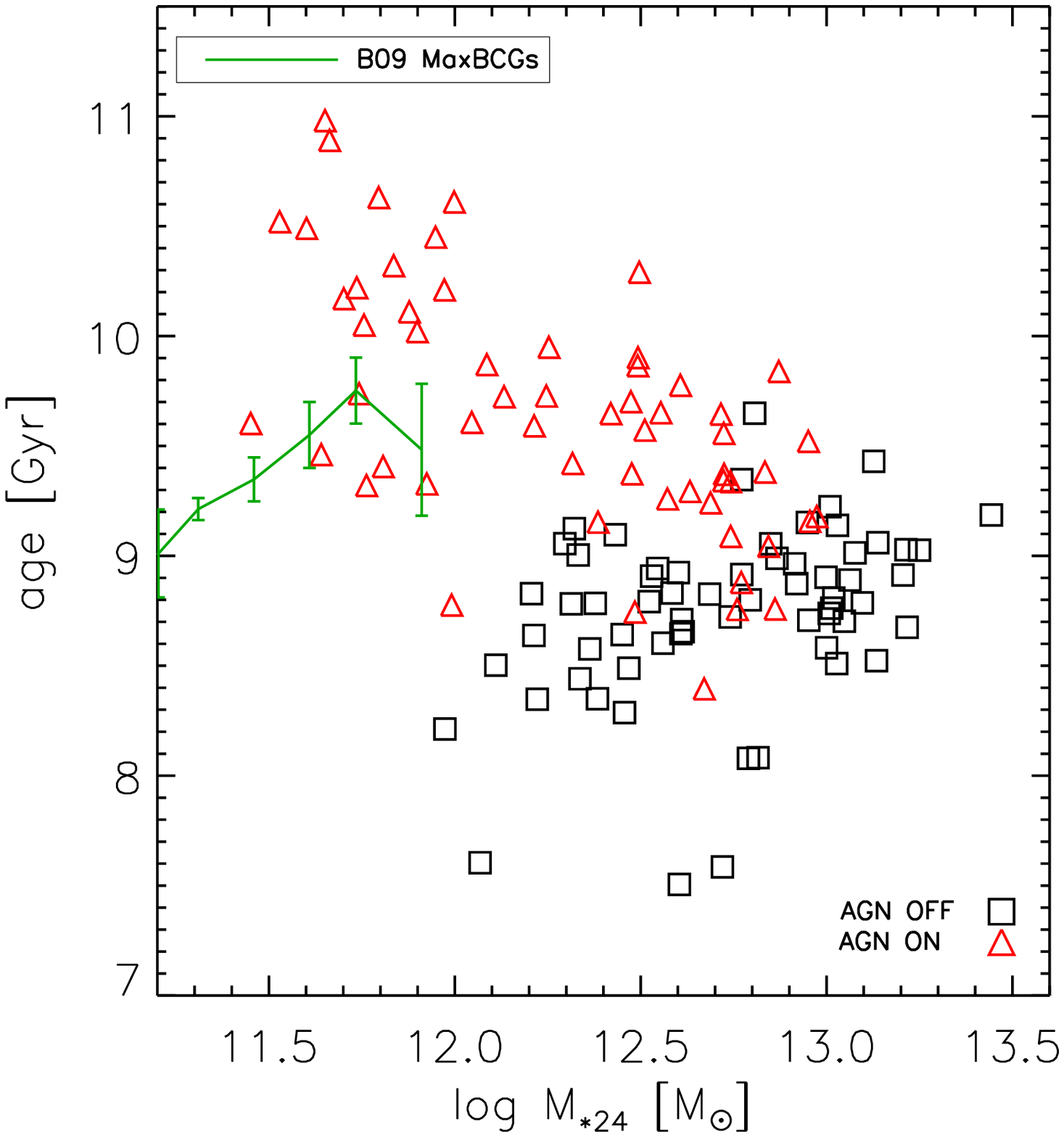}}
  \caption{Mass averaged ages, within 24 mag arcsec$^{-2}$ in the V band,
   of stellar particles as a function of total stellar mass
  in simulated BCGs, compared with observational estimates by Bernardi (2009).}
\label{fig:ages}
\end{figure}

\subsection{The effect of AGN feedback on density profiles}
\label{subsec:prof}

A few papers have pointed out recently that AGN feedback can flatten the
central ($\sim 10$ kpc)  density profiles of both stars and DM in massive
galaxies. {\rev As for stellar profiles,} this has been shown to be the case
either using idealized numerical experiments (Ragone-Figueroa \& Granato 2011;
Martizzi et al.\ 2013) or cosmological zoom-in high resolution simulation of a
few clusters (Sijacki et al.\ 2007; Puchwein et al.\ 2010; Martizzi et al.\
2012a,b). {\rev As for the DM component, it has been shown that the expansion
driven on its density profile by the strong AGN feedback, required to match the
low star formation efficiencies of large DM halos inferred from observations
(e.g.\ Moster et al.\ 2013), more than counteracts the opposite contraction
caused by baryon condensation, both in cosmological as well as in controlled
simulations (Duffy et al.\ 2010; Ragone-Figueroa, Granato \& Abadi 2012). As a
matter of fact, cuspy density profiles of DM in large ETGs, tentatively
inferred from some recent observation (e.g.\ Tortora et al.\ 2010), could be
difficult to reconcile with an strong AGN feedback. }

Our intermediate resolution simulations of a relatively large sample of
clusters confirm this finding. In Fig.\ \ref{fig:perf_bhnew} we show the
density profiles for the BCGs of 4 representative clusters. Comparing the
predicted profiles with and without AGN feedback, it is apparent that this
process has the average effect of flattening the inner ($R \lesssim 30$ kpc)
stellar profiles. The flattening becomes more and more pronounced going toward
the center, hinting to a core starting at $\sim 10-15$ kpc. Although this scale
is only 2-3 times the gravitational softening, and thus about our resolution
limit, Martizzi et al.\ (2012a), with a spatial resolution five times better,
found a core of the same size, which may suggest that we are witnessing to a
real effect, rather than just a numerical artifact. {\rev Even the DM profile
is flattened by the inclusion of AGN feedback by a sizeable amount.}

It is worth noticing that the prediction of cored inner profiles of stars, over
scales of $\sim 10$ kpc, has been regarded as a success of the simulations
including AGN feedback (Martizzi et al.\ 2012a,b), because there have been
reports of cores in observed profiles of cluster elliptical galaxies. However,
the sizes of observed cores are at least one order of magnitude smaller, say
$\lesssim 1$ kpc (see for instance figure 39 in Kormendy et al.\ 2009).
Therefore, a final assessment on the capability of simulations to reproduce the
correct inner density profiles of BCGs would require much higher resolutions
than that of the simulations presented here. Nevertheless, it seems already
established the fact that current models of AGN feedback produce stellar cores,
or significative flattening of stellar profiles, on relatively well resolved
scales, and much larger than observed. Thus, on one hand present day
sub-resolution models of AGN feedback are  not effective enough in diminishing
the global formation of stars in the most massive galaxies, but on the other
hand they seem to be relatively too effective in their centers.

Also, with AGN included, the baryonic component becomes subdominant down to the
most internal regions accessible at our resolution limit. Actually, several
studies point to an important, or even dominant, contribution to the
gravitational field from DM within $R_e$ in bright ETGs (e.g.\ Tortora et al.\
2012; McConnell et al.\ 2012; Grillo 2010).

\subsection{Stellar ages}

In Fig.\ \ref{fig:ages} we show the mass averaged ages of stellar particles as
a function of total stellar mass in simulated BCGs. These are compared with the
observational estimates given by Bernardi (2009).
It is apparent that the inclusion of AGN feedback, besides
reducing the total mass, thus shifting the points leftwards, increases the
average age by 0.5-2 Gyr, moving the points upwards. However, the effect
decreases with increasing mass, resulting in a slight decrease of the age going
to higher masses, a trend not confirmed by observations. This seems to be
another manifestation, together with the excessive stellar masses, of the
progressively insufficient quenching effect of AGN feedback prescription in
massive galaxies.

{\rev Consistently with the less prolonged star forming phase in AGN ON
simulations, the corresponding BCG mass averaged stellar metallicity turns out
to be slightly lower. The predicted metallicity is essentially independent of
the galaxy mass, and is on average about 0.8 and 0.6 solar in AGN OFF and AGN
ON runs respectively.}

\section{Discussion and conclusions}
\label{sec:discussion}

AGN feedback is widely recognized as a required ingredient for simulations, to
avoid the formation of excessively massive galaxies. Indeed, the approximate
inclusion of this effect into our simulated clusters, alleviates the tensions
with observational constraints on the baryonic budget in the various components
of clusters and groups (this work; Planelles et al.\ 2013). In particular, we
have analyzed here the basic properties of BCGs, finding that AGN ON
simulations go in the right direction of reducing by a factor of a few the
final stellar mass of these exceptional objects.

Nevertheless, the results obtained with our implementation of AGN feedback
(similar to that of most cosmological simulations) are only partially
satisfactory. The stellar mass remains still too large by a significant factor,
greater than two and increasing with halo mass, with respect to real BCGs, a
problem which is shared by other numerical works (Fig.\
\ref{fig:m500_mstarmin30}). Also, their basic structural features show some
disagreement with observations. Indeed, the predicted half light radius at a
given stellar mass is larger than observed (Fig.\ \ref{fig:mstar_re}). The AGN
feedback causes a significant flattening of the stellar density profiles. This
flattening increases with decreasing radius, and suggests almost cored profiles
extending to much larger radii than observed (Fig.\ \ref{fig:perf_bhnew}). It
is unlikely that the latter feature is only due to effects of spatial
resolution, since cores of similar sizes ($\sim 10$ kpc) have been observed in
other higher resolution simulations, including AGN feedback (Martizzi et al.\
2012a).

We found that our AGN feedback, while decreasing the final stellar mass by a
factor of a few, affects very little the predicted stellar velocity dispersion
(the inclusion of AGN feedback shifts the points in Fig.\ \ref{fig:faber}
almost horizontally to the left).  It appears therefore unlikely that a more
aggressive choice of parameters, which could reduce the still excessive stellar
mass, would also reduce the velocity dispersions, to better match the observed
values. In order to achieve this, a more fundamental change in the AGN feedback
model seems to be required. A significative decrease of $\sigma$ can be
attained by large scale outflows of gas, as shown by the controlled numerical
experiments by Ragone-Figueroa \& Granato (2011). They followed the dynamical
evolution of a spheroidal stellar system, embedded in a dark matter halo, after
the rapid ejection of large amount of gas. For instance, they found that if the
expelled gas mass is $\sim 40 \%$ of the initial baryon content of the galaxy,
the velocity dispersion of stars can be reduced by $\sim 30\%$ (their figure
4). Moreover, Dubois et al.\ (2013) run cosmological zoom-in simulations for
the formation of six early type galaxies, in which the AGN feedback has been
included in a kinetic rather than thermal form. In principle, the substantial
decrease of velocity dispersion they found, with respect to the no AGN case
(see their figure 18), can arise from two causes: the shallowing of potential
well due to the gas expulsion, and the quenching of the in situ star formation,
which transforms these galaxies into systems built up mostly by accretion and
minor mergers. Indeed, in minor mergers the velocity dispersion can decrease
inversely proportional to the mass, as can be shown by means of simple virial
theorem arguments (e.g.\ Bezanson et al.\ 2009; Naab et al.\ 2009). However,
Ragone-Figueroa \& Granato (2011) demonstrated that the former cause alone can
have an important effect in decreasing the velocity dispersion. Moreover the
 ``in situ quenching'' is, at least in part, present also in simulations with
thermal AGN feedback, which however does not produce by itself important gas
outflows (Barai et al.\, 2013).  In conclusion, outflows appear to be the best
candidates for obtaining values of stellar velocity dispersions closer to
observed ones.

Among the various relationships we considered, when comparing AGN ON with AGN
OFF runs, we found that the former are characterized by a similar or lower
spread. This is particularly evident for the mass-size relations (Fig.\
\ref{fig:mstar_re}), and we have proposed a possible physical explanation,
based on results of our previous work (Ragone-Figueroa \& Granato, 2011).
Conversely, McCarthy et al.\ 2010 found a clearly {\it greater} spread in the
(different) relationships they considered, when AGN feedback was included in
the OWLS simulation. They do not attempt an explanation for that, and it is
impossible for us to assess the origin of their different result. However, we
notice that when we did test runs adopting different, less effective, advection
prescriptions to keep the SMBH at the centers of galaxies (see Appendix
\ref{app:advection}), our decrease of the dispersion in the relationships
obtained from AGN ON runs was largely masked or reversed.

It may be noticed that the dispersions of the $M_{BH}-M_*$ and
$M_{BH}-M_\sigma$ relations produced by the simulations seem significantly
smaller than observed, taking into account the declared observational
uncertainties (Figs.\ \ref{fig:calib}, \ref{fig:mstar24_mbh},
\ref{fig:sigma_mbh}). Unless the latter are underestimated, this suggests a
link between the growth of stellar mass and that of the SMBH less strict than
that produced by the prescriptions we adopt. For instance,  we do not take into
account the evolution of BH spins (e.g. Barausse 2012), which would result in a
distribution of radiative efficiencies rather than a single value, under the
common assumption that they are set by the radius of the last stable orbit
around the BH. The very small dispersions in simulations could also be
contributed by a too effective BH merging, following galaxy merging. Indeed, it
has been shown that merging of galaxies and their host BH tends to strengthen a
preexisting correlation (Peng 2007).

In conclusion, the prescriptions adopted so far in our simulations to describe
AGN feedback, similar if not identical to those of most cosmological
simulations so far, while producing encouraging results especially for the
global properties of clusters (e.g.\ Mccarthy et al.\ 2010; Fabjan et al.\
2010; Planelles et al.\ 2013), seem to require substantial improvements. In
particular, the effectiveness in diminishing the total star formation of most
massive galaxies has to be increased, whilst the relative effectiveness in the
central regions $r\lesssim 10$  kpc should be properly balanced downwards. In
other words, the effect of AGN feedback should be more widely spread over the
galaxy volume, likely by means of large scale gas outflows. These important
outflows, that can be generated by kinetic AGN feedback, should also help in
decreasing the predicted excessive velocity dispersion of BCGs, on which
instead the adopted thermal feedback has little effect.


\section*{Acknowledgements}
The authors would like to thank Volker Springel for making available to us the
non–public version of the GADGET–3 code, Marisa Girardi for useful discussions
and advices, {\rev and the anonymous referee for comments on the manuscript
that improved its quality.} Simulations have been carried out at the CINECA
supercomputing Centre in Bologna, with CPU time assigned through ISCRA
proposals and through an agreement with University of Trieste. We acknowledge
financial support from the European Commission's Framework Programme 7, through
the Marie Curie Initial Training Network CosmoComp (PITN-GA-2009-238356) and
through the International Research Staff Exchange Program LACEGAL. This work
has been supported by the PRIN-INAF09 project "Towards an Italian Network for
Computational Cosmology", by the PRIN-MIUR09 "Tracing the growth of structures
in the Universe", by the PD51 INFN grant, by the Consejo Nacional de
Investigaciones Cient\'{\i}ficas y T\'ecnicas de la Rep\'ublica Argentina
(CONICET) and by the Secretar\'{\i}a de Ciencia y T\'ecnica de la Universidad
Nacional de C\'ordoba - Argentina (SeCyT).


{}

\clearpage

\appendix
\section{AGN feedback model: tests}
\label{AppA}

In this Appendix, we describe in more details the AGN feedback model, and we
show some tests that justify our prescriptions as well as our parameters
choice. To this aim, we use four different simulations of the same Lagrangian
region, centered on a cluster having a mass $M_{200}=3.7 \times
10^{14}$h$^{-1}$ M$_\odot$ at redshift $z=0$. This region is one among those we
use in the main paper, and all simulation details, except for the AGN sector,
are identical to those presented there. For the first simulation we used
without modifications the original BH and AGN feedback model by
\cite{Springel2005b} (A-STD), but with our chosen values for the parameters
describing radiative and feedback efficiencies, and a switch between quasar and
radio modes (Sijacki et al.\ (2007); see Section \ref{app:accre}). The main
problem in using plainly the original scheme A-STD, is that it was thought and
calibrated for non-cosmological high resolution simulations of merging
galaxies, and its use without some modifications leads, in the context of this
paper, to several unwanted and misleading effects, described below. Leaving for
future work the task of improving both the physical model, as well as that of
making its implementation less resolution dependent, our approach here has been
to introduce the minimal changes required to avoid unreasonable results.

In the other three simulations, we modify A-STD introducing one at a time the
three distinctive features of our final model: {\it(i)} we do not allow the BHs
to swallow gas from the simulation, as it would be predicted by their accretion
rates (A-NOSW); {\it (ii)} we use our scheme for BHs advection and mergers
(A-CNTR); and {\it (iii)} we use our prescription for the coupling of the AGN
feedback energy to the gas particles (A-NWEN). Table \ref{table:appendix} lists
our test simulations.

\begin{table*}
  \caption{
    Test simulations for our AGN feedback modified model.
    Column 1: Simulation name;
    Column 2: Model used.
  }
\begin{tabular}{c || c ||}
\hline
A-STD & \cite{Springel2005b} model + quasar/radio mode\\
A-NOSW & As A-STD, but no gas removal from simulation\\
A-CNTR & As A-STD, but with our modification to the BH advection and merger
schemes\\
A-NWEN & As A-STD, but with our new coupling between AGB feedback
energy and gas particles\\
\hline
\end{tabular}
\label{table:appendix}
\end{table*}

\subsection{BH accretion and ensuing feedback}
\label{app:accre}

We identify DM haloes using an on-the-fly Friends-of-Friends algorithm. We seed
a BH particle every time that a DM halo has a mass $M_{halo}>M_{th}=2.5 \times
10^{11} h^{-1}$ and does not already contain a BH. We initially put the newly
seeded BH at the position and with the velocity of the gas particle having the
minimum gravitational potential energy, and we remove the latter from the
simulation.  The initial mass of the BH is set to $M_{seed}=5 \times 10^6
h^{-1}$ M$_\odot$, but see Section \ref{app:swallow}.

As in most previous simulations with AGN feedback, also in our simulations the
BH grows at a rate which is the minimum between what we dub {\it
$\alpha$-modified} Bondi rate \citep{Bondi52} and the Eddington rate:
\begin{equation}
\dot{M}_{BH} = \min\left(\dot{M}_{Bondi,\alpha},\dot{M}_{Edd}\right)
\label{eq:accretion}
\end{equation}
The general idea is that the former should provide an order of magnitude
estimate of the gas which potentially can feed the BH, given its physical
conditions, without producing enough luminosity to reverse the flow, as
controlled by the latter. Of course, we should remind that this is an
over-simplification of a very complex process, and that these rates are
obtained under several strong assumptions. As a result, they could even fail to
provide a correct order of magnitude.

\begin{figure}
\hspace{-1cm}
 \includegraphics[width=9.cm, height=8.5cm]{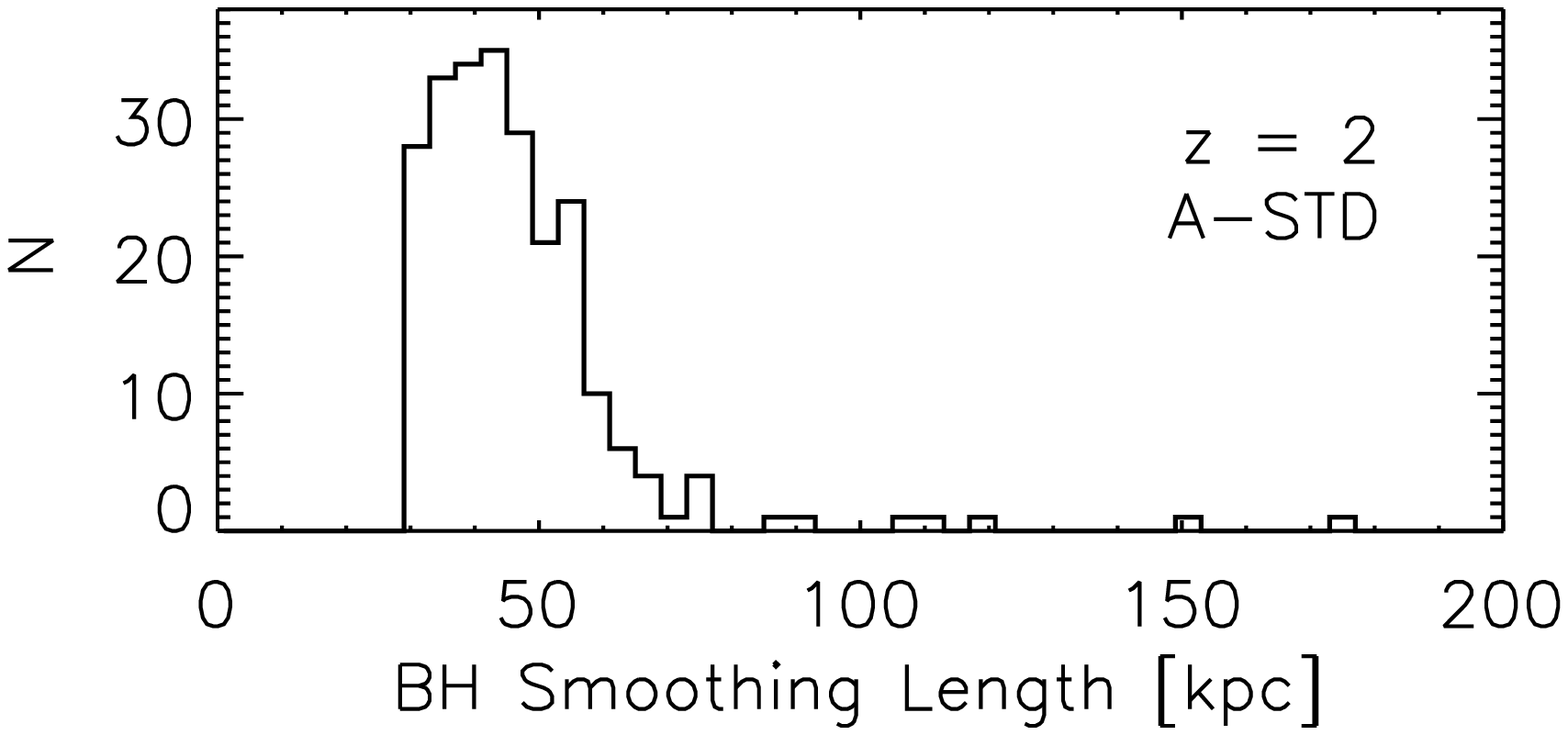}
 \vspace{-4.5cm}
 \caption{Histogram of the BH particles smoothing length in the test region, for the
 A-STD simulation at redshift 2. This is representative of the situation in the redshift
 range most relevant for BH activity.}
 \label{fig:smolen}
\end{figure}

The $\alpha$-modified Bondi rate is
\begin{equation}
  \dot{M}_{Bondi,\alpha} = \alpha \, 4 \pi  \frac{G^2M^2_{BH}}{c^2_s+v^2_{BH}}\rho
\label{eq:bondi}
\end{equation}
where $M_{BH}$ is the BH mass, $v_{BH}$ its velocity (relative to the
surrounding gas bulk motion), $c_s$ the sound velocity of the gas surrounding
the BH and $\rho$ its density,  and a fudge adimensional factor $\alpha\sim
100$ is included. This apparently arbitrary and large factor, has been
initially justified simply by the pragmatic requirement of helping in producing
a reasonable black hole mass at the end of the simulation. More recently, it
has been perceived as a cure for the limitation of having to use gas properties
numerically calculated on scales much larger than the BH sphere of influence.
Indeed, we always estimate gas quantities using the SPH spline kernel, whose
smoothing length is the radius of a sphere containing a mass of $4 \times
N_{neigh} M_{gas,init}$ ($N_{neigh}$ is the number of neighbours for hydro
calculations and $M_{gas,init}$ the initial mass of gas particles). This scale
length is typically $\gtrsim 10$ kpc in cosmological simulation (see e.g.\
Fig.\ \ref{fig:smolen}), thus is is much larger than the real scale over which
the BH accretes its gas (as discussed at the end of Section \ref{app:swallow}).
The assumption that gas properties at scales as large as these are related to
the corresponding small-scale is unjustified. Actually, Booth \& Schaye (2009)
pointed out that the un-sufficient resolutions underestimate the density and
overestimate the sound speed by orders of magnitude, thus justifying large
values of $\alpha$. However, this would suggest some relationship between
$\alpha$ and the resolution.

The Eddington accretion rate is:
\begin{equation}
\dot{M}_{Edd} = 4 \pi \frac{c \, G \, m_p}{\epsilon_r c^2 \sigma_T} M_{BH}
\label{eq:edd}
\end{equation}
where $m_p$ is the proton mass, $\sigma_T$ the Thompson cross section, $c$ the
speed of light and $\epsilon_r$ the radiative efficiency, giving the radiated
energy in units of the energy associated to the accreted mass. Once we estimate
the accretion rate, we assume that a fraction $\epsilon_f$ of the irradiated
energy couples to the IGM/ICM gas and heats it. Thus the rate of available
feedback energy from an AGN is:
\begin{equation}
\dot{E}_{feed} = \epsilon_f \epsilon_r \dot{M}_{BH}c^2
\label{eq:feedback}
\end{equation}

Finally, following Sijacki et al.\ (2007) we assume a transition from a {\it
quasar mode} to a {\it radio mode} of AGN feedback. This occurs when the
accretion rate becomes smaller than a given limit, $\dot{M}_{BH}/\dot{M}_{Edd}
= 10^{-2}$. In this case, we increase the feedback efficiency $\epsilon_f$ by a
factor four.

We varied the parameters $\alpha$, $\epsilon_r$ and $\epsilon_f$  to match the
BH mass- stellar mass relation (see Figs. \ref{fig:calib} and
\ref{fig:mstar24_mbh}). Our choice is $\alpha=100$, $\epsilon_r=0.2$ and
$\epsilon_f=0.2$.

\subsection{Numerical aspects of BH gas accretion}
\label{app:swallow} The growth rate of SMBHs is calculated as described above.
In literature, a scheme for removing gas from the simulation following as much
as possible the BH mass accretion rate (''swallowing'' of gas particle by BHs)
is usually adopted, to ensure mass conservation. Commonly, gas particles within
the smoothing length of BHs are randomly removed from the simulation, with a
probability proportional to the SPH kernel, and their mass is added to the BH
mass. Of course this leads to a very discretized representation of the process,
and the problem is particularly severe in cosmological simulations, wherein the
mass of a gas particles ($\sim 10^8 M_\odot$ in our case) can be in the range
of an already well developed SMBH. Also, the swallowing occurs often at
distances orders of magnitude greater than the actual sphere of influence of
the BH.

To cope with this, the models usually characterize the BH particles by means of
{\it two} masses: a {\it theoretical} one $M_{BH,th}$, determined exactly by
the BH accretion rate computed from the Eq.\ \ref{eq:accretion} and used for
all the computations involving AGN feedback, and a {\it dynamical} BH mass
$M_{BH,dyn}$, used for the gravitational and dynamical calculations. The latter
is adjusted by swallowing gas particles from time to time, in order to keep it
as close as possible to the theoretical mass. At seeding, the BH is created by
converting a gas particle into a BH particle, with the same dynamical mass.

However, at the relatively low mass resolution of cosmological simulations like
our own, the match between the two BH masses is quite unsatisfactory until the
SMBH reaches a very high mass $\gtrsim 10^9 M_\odot$. For instance, at seeding,
the BH begins its  ``life'' with a dynamical mass orders of magnitude greater than
the theoretical one. On the other hand, a later seeding, with a theoretical
mass already close to the gas particle mass, would miss the feedback effects of
the BH during an important part of its growth. In cosmological simulations, the
large ratio $M_{BH,dyn}/M_{BH,th}$, which should approach unity while the BH
grows by accretion, thanks to the algorithm of swallowing, can instead persist
for BH particles of large theoretical masses, if their mass growth occurs
mostly by mergers. This situation can occur particularly often when the
numerical scheme leads to some {\it overmerging} (see Section
\ref{app:advection}). In this case, the dynamical mass of the particle, used to
compute its gravitational effects, can become really huge.

To avoid introducing artifacts due to the large difference between the two
masses, we choose here to keep always the dynamical mass equal to the
theoretical one, as given by the accretion formula Eq.\ \ref{eq:accretion}.
Given that this already means to give up to rigorous mass conservation, we also
decided to avoid at all any swallowing of the gas particles. The resulting
global error in term of mass non-conservation is small, of the order of a
fraction of percent the stellar mass of the halo where the BH dwells, and can
be neglected on galactic scales. The advantage is that it avoids the gas
depletion on physical scales much larger than the real sphere of influence of
SMBH.

Indeed, typical values of BH smoothing lengths are of the order of tens of kpc
in the redshift range more relevant for BH accretion. An example for z=2 is
shown in Fig.\ \ref{fig:smolen}. By converse, the radius at which a central
SMBH dominates the potential of the host galaxies is
\begin{equation}
r_{BH} \simeq 11 \frac{M_{BH}}{10^8 M_\odot} \left(\frac{\sigma}{200 \, \mbox{km/s}}\right)^{-2} \mbox{pc}
\end{equation}
or, assuming that the system obeys the observed correlation between stellar
velocity dispersion $\sigma$ and BH mass,
\begin{equation}
r_{BH} \simeq 16 \left(\frac{M_{BH}}{10^8 M_\odot}\right)^{0.62} \mbox{pc}
\end{equation}
where we have used the recent determination by McConnell \& Ma (2013) for ETGs.

In particular, we found that this unrealistic removal of the gas has, among
others, also the effect of un-physically shallowing the gravitational potential
in the innermost part of the haloes, making it easier for the BHs to drift away
from it.

\subsection{BH advection and mergers}
\label{app:advection}

The effect of the feedback energy provided by SMBHs critically depends on the
fact that particles representing them stay at the center of their parent DM
haloes, where it is seeded (see e.g.\ \citet{Wurster13}). In addition, if a BH
un-physically exits from its parent DM halo, that halo becomes eligible for
seeding another spurious BH. This stability is not attained by construction in
large-scale numerical simulations, for two reasons. The first is that, when the
BH has a dynamical mass still comparable to that of other numerical particles,
it can be scattered around by  two-body encounters. The second has to do with
the chaotic nature of self-gravitating systems is. The trajectory of an
individual numerical particle in phase-space is subject, due to numerical
errors, to large deviations from the exact trajectory of a particle starting
from the same initial conditions. This effect  is particularly important for
un-softened systems (see e.g. Quinlan \& Tremaine 1992), but it cannot be
ignored also when a softened gravitational force is used (see e.g. Hayes 2003).
Therefore it is necessary to introduce an artificial advection algorithm to
keep the SMBH as close as possible to the galactic center. In our case, at each
time step of the BH particle, we force it at the position of the most bound
particle, of whatever type, within the gravitational softening of the BH.

When two BHs are within the gravitational softening and their relative velocity
is smaller than a fraction of the sound velocity of the surrounding gas, we
merge them. We place the resulting BH at the position of the most massive one.
This repositioning scheme clearly has the consequence of violating the momentum
conservation; however, the aim of our numerical prescriptions is to reproduce
some physical properties and effects emerging from unresolved scales, and  not
to directly simulate the BH physics. From this point of view, a limited
non-conservation of momentum (or mass, Section \ref{app:swallow}) is
acceptable.

\begin{figure}
 \centerline{\includegraphics[width=8.5cm, height=8.5cm]{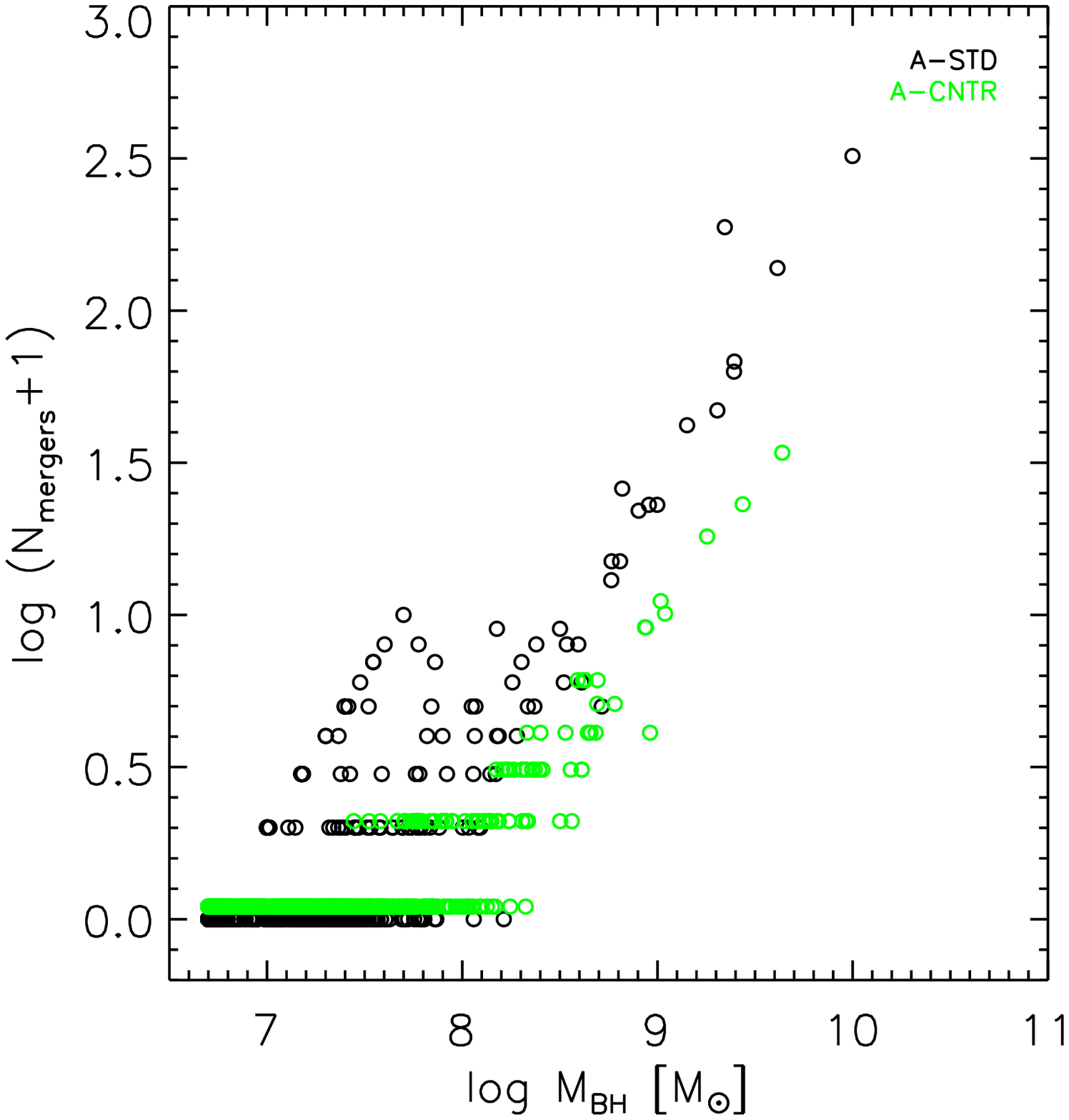}}
 \caption{Total number of mergers of the BHs in the test region (down to z=0), against their
 mass, for the A-STD simulation, closely following the Springel et al.\ (2005) scheme,
 and the A-CTRL, in which instead our modifications for the advection and merging algorithms have been
 introduced.}
 \label{fig:BH_n_merge}
\end{figure}

\begin{figure}
\hspace{-1cm}
 \includegraphics[width=9.cm, height=8.5cm]{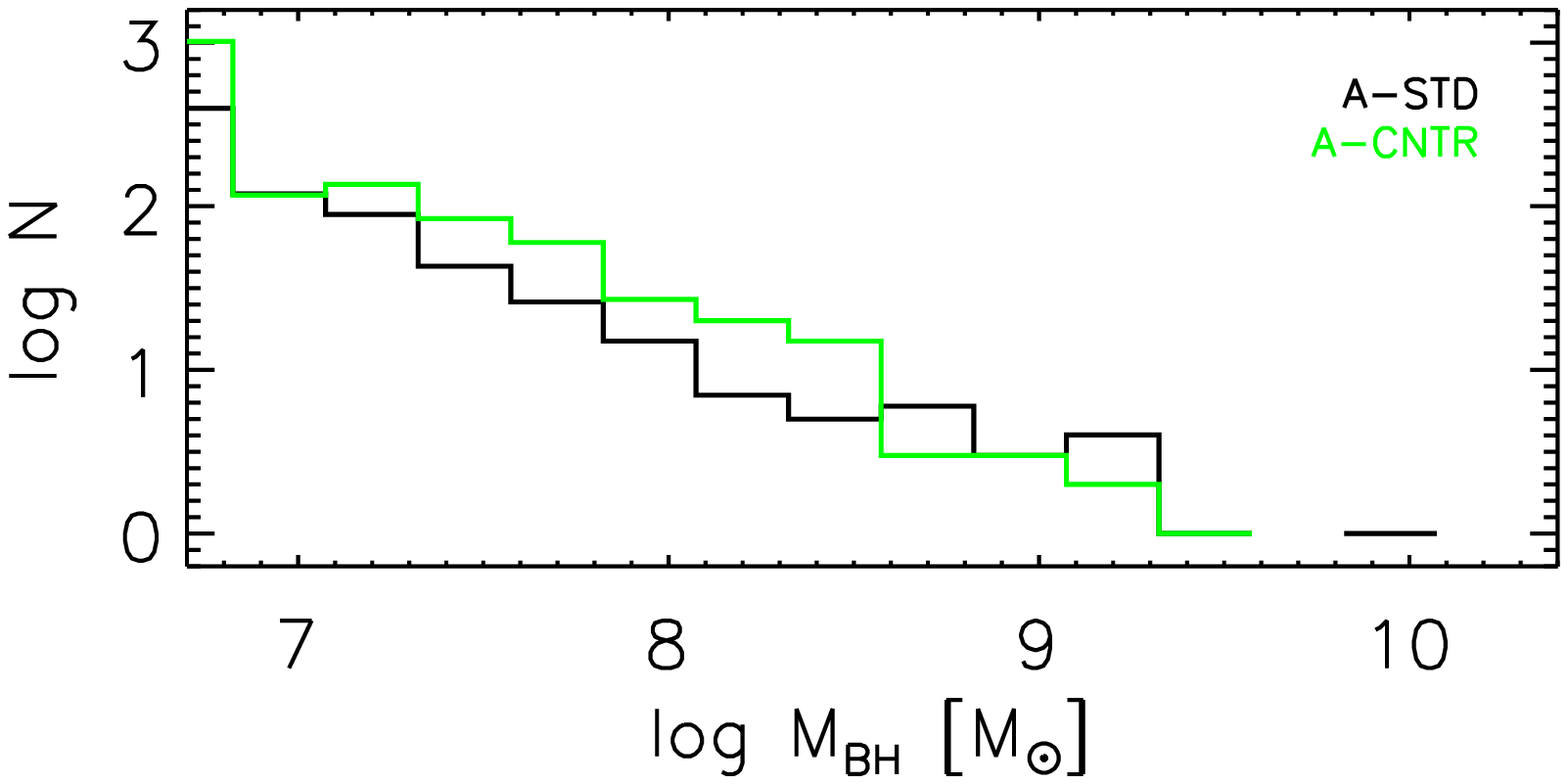}
 \vspace{-4.5cm}
 \caption{
   Mass distribution of BH at redshift $z=0$.}
 \label{fig:BHmassfctn}
\end{figure}

We explicitly note that we use the gravitational softening as a searching
length, rather than the BH smoothing length as in most previous
implementations, for both the advection and the merging algorithms. In
principle, the former seems to be a more sensitive choice, because the
numerical processes acting to drift away the BH, as well as the physical
process leading to BH merging, are gravitational in nature. Furthermore, at the
resolution of our simulations, the BH smoothing length is unreasonably large
for these purposes, and significantly larger than the gravitational softening
(Fig. \ref{fig:smolen}). In particular, if we used the former for searching the
minimum potential particle, BHs dwelling in satellite haloes could often be
spuriously displaced to the center of a more massive DM halo, and immediately
would merge with the BH dwelling there. This numerical over-merging affects the
mass function of BH, artificially increasing by merging the final masses at the
high mass end, and depleting the number of BH at low and intermediate masses.
It is significantly reduced by repositioning within the softening length, as
can be seen from Figures \ref{fig:BH_n_merge} and \ref{fig:BHmassfctn}. For
example, the most massive BH had 322 mergers in A-STD, which closely follows
the Springel et al.\ (2005) scheme, but only 34 in A-CNTR, wherein our
modifications for the advection and merging algorithms have been introduced. As
a result, in the latter case its final mass is twice smaller.

\subsection{Feedback energy distribution}
\label{app:ene}

In the original scheme by \cite{Springel2005b}, the AGN feedback energy is
simply added to the specific internal energy of gas particles. Owing to the
features of the effective model of star formation and stellar feedback
\citep{SH03}, when SMBH energy is given to a star-forming gas particle, it is
almost completely lost. This is due to the fact that any deviation of the
internal energy from the equilibrium one rapidly decays, over a timescale
shorter than a typical timestep.

Indeed, the total specific equilibrium energy is given by $u = (\rho_h u_h +
\rho_c u_c) \rho^{-1}$, where $u_h$ ($\rho_h$) and $u_c$ ($\rho_c$) are the
equilibrium specific energies (densities) of the hot and cold phase
respectively of the star-forming gas. $u_c$ is assumed to be constant, while
\begin{equation}
    u_h = \frac{u_{SN}}{A+1}+u_c
\end{equation}
where $A$ is the efficiency of evaporation of cold clouds in the Inter-Galactic
Medium (IGM), taken to be $A \propto \rho^{-4/5}$; $u_{SN}$ is the specific
energy provided by supernovae, also assumed constant \citep{SH03}.

From the same work, the time-scale over which deviations from this equilibrium
energy decay is:
\begin{equation}
    \tau_h = \frac{t_* \rho_h}{\beta (A+1) \rho_c}
\end{equation}
where $t_* \propto \rho^{-1/2}$ is the star formation time-scale, $\rho$ being
the average gas particle density; $\beta$ the mass fraction of stars which
immediately explode as Type II supernovae. In particular, deviations due to an
external source (external from the point of view of the effective model), such
as AGN feedback evolves over a time-step $\Delta t$ to $u + (u^{out}-u)
e^{-\Delta t/\tau_h}$, where $u^{out}$ is the energy provided from outside. In
star-forming region, the higher the density of the gas is, the faster is the
convergence to the equilibrium specific energy $u$ and the least effective is
the AGN energy feedback. Any increase of internal specific energy of
star-forming particles is immediately lost if the effective model for star
formation and feedback is active.

To avoid this, whenever a star-forming gas particle receives energy from a
SMBH, we now calculate the temperature $T_{c}$ at which the  cold gas phase
would be heated by it. The AGN energy is given to hot and cold phases
proportionally to their mass, but since the mass fraction of the cold phase is
high, this particular choice is not important. If $T_{c}$ turns out to be
larger than the average temperature of the gas particle (before receiving AGN
energy), we consider the particle not to be multi-phase anymore and prevent it
from forming star. To avoid an immediate re-entering of this gas particle in
the multi-phase star forming state, wherein it would immediately lose all of
its internal specific energy in excess to the equilibrium one, we also add a
maximum temperature condition to the usual minimum density condition for a gas
particle to become multi-phase and form stars.
\begin{figure}
 \centerline{\includegraphics[width=8.5cm, height=8.5cm]{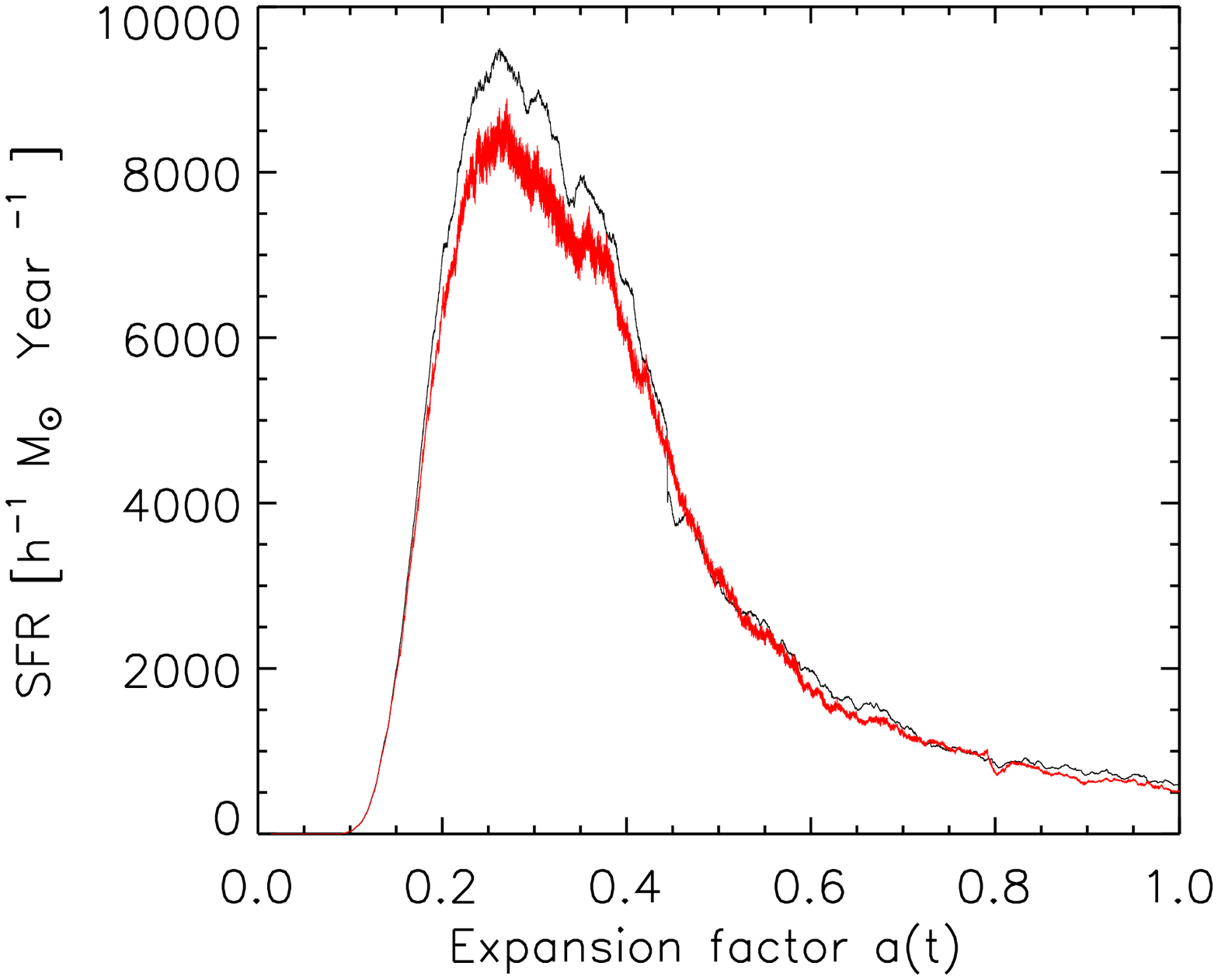}}
 \caption{
   Star formation rate in our selected Lagrangian regions for
   simulation A-STD (black line) and A-NWEN (red line)}
 \label{fig:BHSFR}
\end{figure}

\begin{figure}
 \centerline{\includegraphics[width=8.5cm, height=8.5cm]{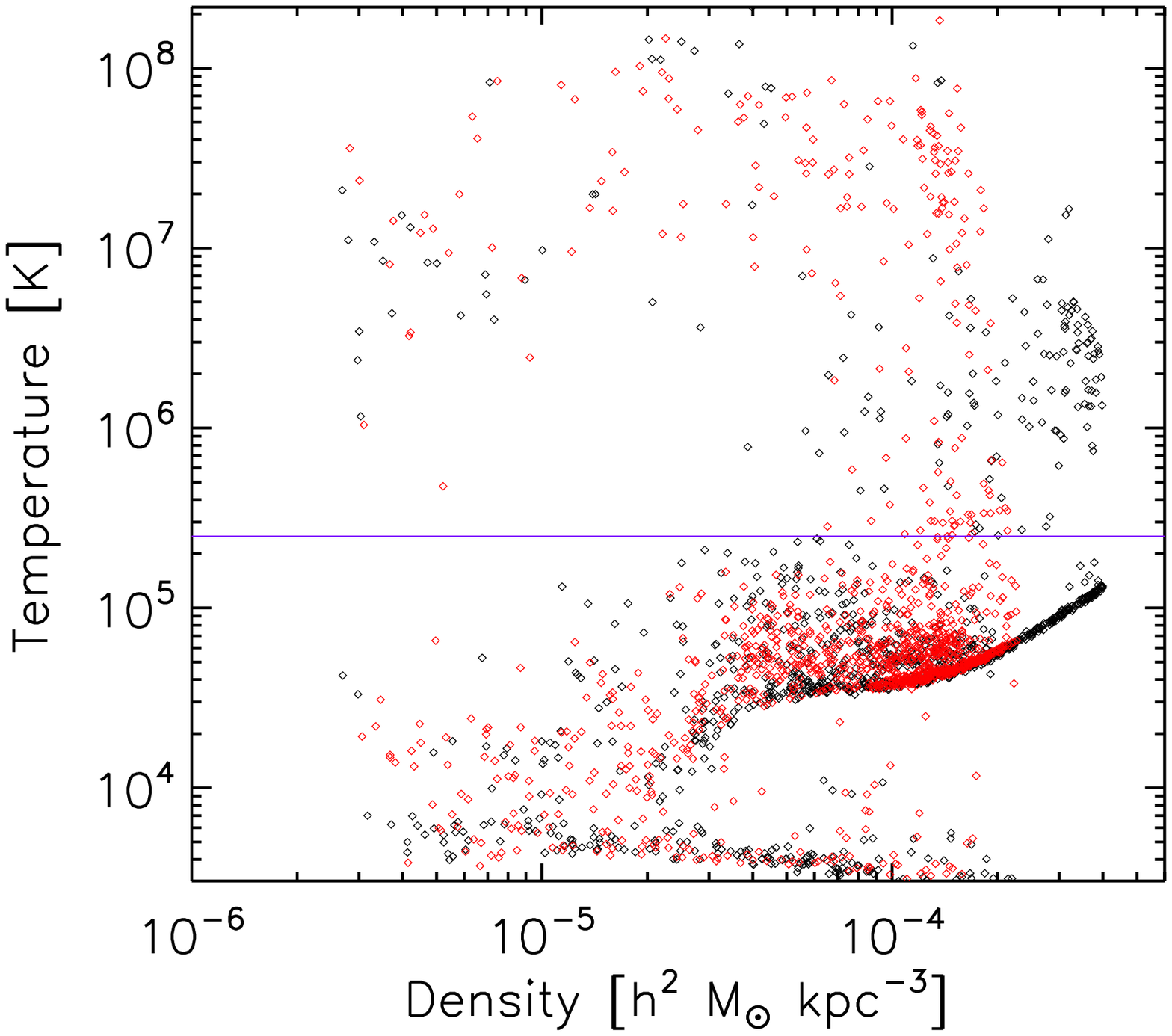}}
 \caption{
   Phase diagram for gas particles in a sphere of radius
   $R=50$h$^{-1}$ comoving kpc, centered on the most massive BH at
   redshift $z=2.8$. Black diamonds: simulation A-STD; red diamonds:
   simulation A-NWEN. The horizontal line marks our adopted temperature
 threshold for a particle to become multi-phase and star-forming.}
 \label{fig:BHphase}
\end{figure}

In Fig. \ref{fig:BHSFR}, we show the effect of our new prescription on the
overall star formation rate comparing simulation A-STD (black line) with A-NWEN
(red line). A reduction of $\sim 10 \%$ is appreciable at the peak of the star
formation, which happens between redshift $z=3$ and $z=2.5$. The reason for
this additional quenching can be understood from Fig. \ref{fig:BHphase}, where
we show the density-temperature diagram of gas particles in simulations A-STD
(black diamonds) and A-NWEN (red diamonds). We consider all the particles
included in a sphere of radius $R=50$ h$^{-1}$ comoving kpc, centered on the
most massive BH at redshift $z=2.8$. With the original prescription, even in
presence of AGN feedback, gas particles remain very dense and star forming: the
bulk of them stays on the effective equation of state given by the star
formation model. This is tracked by the line shaped concentration of points in
the lower right corner of the diagram. This is caused by the quick convergence
to the equilibrium specific energy explained above. With our new scheme, gas is
less dense, and more particles can reach temperatures higher than $10^7$K when
receiving feedback energy, becoming non star forming. Thus, with our scheme the
feedback energy acts in a more efficient way.

Finally, we also found that, contrarily to naive expectation, in the original
model an increase of AGN feedback efficiency $\epsilon_f$ gives an increase in
the star formation rate, rather than a reduction. This disturbing feature is
due to the increase of gas pressure of star forming particles, caused by the
action of {\it non} star-forming particles, heated by feedback, on the formers.
Indeed, in the effective model of \cite{SH03}, an increase of gas pressure
enhances the star formation. With our new scheme, this effect is reduced, since
the pressure increase is partly balanced by a density decrease, as shown in
Fig. \ref{fig:BHphase}. However it is still not eliminated. A substantial
refinement of the AGN feedback prescriptions which pays particular attention to
the interplay between AGN energy and star formation model is still needed in
cosmological simulations.

\end{document}